\documentclass[preprint]{imsart}
\pdfoutput=1 

\RequirePackage[OT1]{fontenc}
\RequirePackage{amsthm,amsmath}
\RequirePackage[numbers]{natbib}
\RequirePackage[colorlinks,citecolor=blue,urlcolor=blue]{hyperref}
\RequirePackage{fancyvrb}
\RequirePackage{alltt}
\RequirePackage{geometry}
\RequirePackage{booktabs}
\RequirePackage{microtype}

\RequirePackage[scaled=0.92]{helvet}
\RequirePackage{palatino,mathpazo}
\RequirePackage[scaled=1.02]{inconsolata}
\RequirePackage[T1]{fontenc}

\usepackage{graphicx}
\usepackage{float}
\usepackage{amssymb}
\usepackage{multicol}
\allowdisplaybreaks
\arxiv{arXiv:0000.0000}

\startlocaldefs
\numberwithin{equation}{section}
\theoremstyle{plain}

\newtheorem{algorithm}{Algorithm}
\newcommand {\bs} { \boldsymbol}
\DeclareRobustCommand\code{\bgroup\@noligs\@codex}
\def\@codex#1{\texorpdfstring%
{{\normalfont\ttfamily\hyphenchar\font=-1 #1}}%
{#1}\egroup}

\DeclareRobustCommand\samp{`\bgroup\@noligs\@sampx}
\def\@sampx#1{{\normalfont\texttt{#1}}\egroup'}

\newcommand{\file}[1]{{`\normalfont\textsf{#1}'}}

\let\pkg=\strong
\newcommand{\CRANpkg}[1]{\href{https://CRAN.R-project.org/package=#1}{\pkg{#1}}}%

\DefineVerbatimEnvironment{example}{Verbatim}{}
\renewenvironment{example*}{\begin{alltt}}{\end{alltt}}

\DefineVerbatimEnvironment{Sinput}{Verbatim}{}
\DefineVerbatimEnvironment{Soutput}{Verbatim}{}
\DefineVerbatimEnvironment{Scode}{Verbatim}{}
\DefineVerbatimEnvironment{Sin}{Verbatim}{}
\DefineVerbatimEnvironment{Sout}{Verbatim}{}

\endlocaldefs

\begin{document}

\begin{frontmatter}
\title{\CRANpkg{BayesBinMix}: an R Package for Model Based Clustering of Multivariate Binary Data}
\runtitle{BayesBinMix}
\thankstext{T1}{{\tt  https://CRAN.R-project.org/package=BayesBinMix}}

\begin{aug}
\author{\fnms{Panagiotis} \snm{Papastamoulis}} \and
\author{\fnms{Magnus} \snm{Rattray}}

\affiliation{University of Manchester}

\address{
Faculty of Life Science\\
Michael Smith Building\\
Oxford Road\\
Manchester, UK\\
{\tt panagiotis.papastamoulis@manchester.ac.uk}\\
{\tt magnus.rattray@manchester.ac.uk}}
\end{aug}

\begin{abstract}
The \pkg{BayesBinMix} package offers a Bayesian framework for clustering binary data with or without missing values by fitting mixtures of multivariate Bernoulli distributions with an unknown number of components. It allows the joint estimation of the number of clusters and model parameters using Markov chain Monte Carlo sampling. Heated chains are run in parallel and accelerate the convergence to the target posterior distribution. Identifiability issues are addressed by implementing label switching algorithms. The package is demonstrated and benchmarked against the Expectation-Maximization algorithm using a simulation study  as well as a real dataset.
\end{abstract}

\begin{keyword}
\kwd{CRAN package}
\kwd{mixture models}
\kwd{clustering binary data}
\kwd{trans-dimensional MCMC}
\kwd{Metropolis-coupled MCMC}
\end{keyword}

\end{frontmatter}

\section{Introduction}

Clustering data is a fundamental task in a wide range of applications and finite mixture models are widely used for this purpose \citep{McLachlan:00, Marin:05, Fruhwirth:06}. In this paper our attention is focused on clustering binary datasets. A variety of studies aims at identifying patterns in binary data including, but not limited to, voting data \citep{ilin2012unsupervised}, text classification \citep{juan2002use}, handwritten digit recognition \citep{al2003databases}, medical research \citep{BIOM:BIOM762}, animal classification \citet{li2005general} and genetics \citep{abel1993autologistic}. 

Throughout this paper the term ``cluster'' is used as a synonym of ``mixture component''. Finite mixture models can be estimated under a frequentist approach using the Expectation-Maximization (EM) algorithm \citep{Dempster:77}. However, the likelihood surface of a mixture model can exhibit many local maxima and it is well known that the EM algorithm may fail to converge to the main mode if it is initialized from a point close to a minor mode. Moreover, under a frequentist approach, the selection of the number of clusters is not straightforward: a mixture model for each possible value of number of clusters is fitted and then the optimal one is selected according to penalized likelihood criteria such as the Bayesian information criterion \citep{Schwarz:78} or the Integrated complete likelihood criterion \citep{Biernacki:2000}. The reader is also referred to \citet{lindsay1995mixture, bohning2000computer} for non-parametric likelihood estimation of a mixture model.

On the other hand, the Bayesian framework allows to put a prior distribution on both the number of clusters as well as the model parameters and then (approximately) sample from the joint posterior distribution using Markov chain Monte Carlo (MCMC) algorithms \citep{Richardson:97, stephens2000, Nobile2007, white2016bayesian}. However this does not mean that the Bayesian approach is not problematic. In general, vanilla MCMC algorithms may require a very large number of iterations to discover the high posterior density areas and/or sufficiently explore the posterior surface due to the existence of minor modes. Second, identifiability issues arise due to the label switching phenomenon \citep{redner1984mixture} which complicate the inference procedure. 

The \pkg{BayesBinMix} package explicitly takes care of the previously mentioned problems for the problem of clustering multivariate binary data: 
\begin{enumerate}
\item Allows missing values in the observed data
\item Performs MCMC sampling for estimating the posterior distribution of the number of clusters and model parameters
\item Produces a rapidly mixing MCMC sample by running parallel heated chains which can switch states
\item Post-processes the generated MCMC sample and produces meaningful posterior mean estimates using state of the art algorithms to deal with label switching.
\end{enumerate} 

The rest of the paper is organised as follows. The mixture model is presented in Section \ref{sec:model}. Its prior assumptions and the corresponding hierarchical model is introduced in Section \ref{sec:prior}. The  basic MCMC scheme is detailed in Section \ref{sec:sampler}. Section \ref{sec:labelSwitching} deals with post-processing the generated MCMC sample in order to overcome identifiability issues due to the label switching problem. Finally, the basic sampler is embedded in a Metropolis-coupled MCMC algorithm as described in Section \ref{sec:mc3}. The main function of the package is described in Section \ref{sec:cmFunction}. Simulated and real datasets are analyzed in Sections \ref{sec:sim} and \ref{sec:realdata}, respectively.

\section{Model}\label{sec:model}

Let $\bs x =(x_1,\ldots,x_n)$ denote a random sample of multivariate binary data, where $x_i = (x_{i1},\ldots,x_{id})$; $d > 1$, for $i = 1,\ldots,n$. Assume that the observed data has been generated from 
a mixture of independent Bernoulli distributions, that is,
\begin{alignat}{1}\nonumber
x_i&\sim\sum_{k=1}^{K} p_k\prod_{j = 1}^{d}f\left(x_{ij};\theta_{kj}\right)\\
& =  \sum_{k=1}^{K} p_k\prod_{j = 1}^{d}\theta_{kj}^{x_{ij}}\left(1-\theta_{kj}\right)^{1-x_{ij}}\mathbb{I}_{\{0,1\}}(x_{ij})\label{eq:mixture},
\end{alignat}
independently for $i = 1,\ldots,n$, where $\theta_{kj}\in\Theta=(0,1)$ denotes the probability of success for the $k$-th cluster and $j$-th response for $k = 1,\ldots,K$; $j = 1,\ldots,d$,  $\bs p = (p_1,\
\ldots,p_K)\in\mathcal P_{K-1}=\{p_k;k = 1,\ldots,K-1: 0\leqslant p_k \leqslant 1;0\leqslant p_K = 1-\sum_{k=1}^{K-1}p_k\}$ corresponds to the vector of mixture weights and $\mathbb{I}_{A}(\cdot)$ denotes the indicator function of a (measurable) subset $A$. 

It is straightforward to prove that the variance-covariance matrix of a mixture of independent Bernoulli distributions is not diagonal (see e.g.~\cite{bishop}), which is the case for a collection of independent Bernoulli distributions. Therefore, the mixture model exhibits richer covariance structure thus it can prove useful to discover correlations in heterogeneous multivariate binary data.

The observed likelihood of the model is written as
\begin{equation}\label{eq:observedL}
L_K(\bs p,\bs\theta;\bs x) = \prod_{i=1}^{n}\sum_{k=1}^{K} p_k\prod_{j = 1}^{d}\theta_{kj}^{x_{ij}}\left(1-\theta_{kj}\right)^{1-x_{ij}}, \quad (\bs p,\bs\theta)\in\mathcal P_{K-1}\times\Theta^{Kd}
\end{equation}
where $\bs x\in \mathcal X^{n} = \{0,1\}^{nd}$.
For any fixed value of $K$, Equation \eqref{eq:observedL} can be further decomposed by considering that observation $i$ has been generated from the $z_i$-th mixture component, that is, 
\begin{equation}\label{eq:XgivenZ}
x_i|z_i = k \sim \prod_{j=1}^{d}f\left(x_{ij};\theta_{kj}\right),\quad{\mbox{independent for } i = 1,\ldots,n}.\end{equation}
Note that the allocation variables $z_i\in\mathcal Z_K = \{1,\ldots,K\}$; $i = 1,\ldots,n$ are unobserved, so they are treated as missing data. Assume that \begin{equation}
P\left(z_i = k|\bs p,K\right) = p_k,\quad k = 1,\ldots,K \label{eq:priorZ}
\end{equation} 
and furthermore that $(x_i,z_i)$ are independent for $i = 1,\ldots,n$. Data augmentation \citep{tanner} considers jointly the \textit{complete data}  $\{(x_i,z_i);i=1,\ldots,n\}$ and it is a standard technique exploited both by the Expectation-Maximization algorithm \citep{Dempster:77} as well as the Gibbs sampler \citep{gelfand}. The \textit{complete likelihood} is defined as
\begin{alignat}{1}\nonumber
L_K^{c}\left(\bs p,\bs\theta;\bs x, \bs z\right) &= \prod_{i=1}^{n}p_{z_i}\prod_{j=1}^{d}\theta_{z_{i}j}^{x_{ij}}\left(1-\theta_{z_{i}j}\right)^{1-x_{ij}}\\
&=\prod_{k=1}^{K}p_k^{n_k}\prod_{j=1}^{d}\theta_{kj}^{s_{kj}}\left(1-\theta_{kj}\right)^{n_k - s_{kj}},\quad (\bs p,\bs\theta)\in\mathcal P_{K-1}\times\Theta^{Kd}\label{eq:completeL}
\end{alignat}
where $n_k = \sum_{i=1}^{n}\mathbb I(z_i = k)$ and $s_{kj} = \sum_{i=1}^{n}\mathbb I(z_i = k)x_{ij}$, $k =1,\ldots,K$; $j = 1,\ldots,d$, for a given $(\bs x, \bs z)\in \mathcal X^{n}\times\mathcal Z^{n}_K$.

\section{Prior assumptions}\label{sec:prior}

Note that the quantities $\bs p,\bs\theta,\bs z$ are defined conditionally on $K$. For convenience we will assume that $K\in\mathcal K = \{1,\ldots,K_{\mbox{max}}\}$, where $K_{\mbox{max}}$ denotes an upper bound on the number of clusters. Hence, under a model-based clustering point of view, the vector $$(K,\bs p,\bs\theta,\bs z)\in\mathcal A := \mathcal K\times\mathcal P_{K-1}\times\Theta^{Kd}\times\mathcal Z_K^{n}$$ summarizes all unknown parameters that we wish to infer.

The following prior assumptions are imposed
\begin{alignat}{1}\label{eq:priorK}
K &\sim \mbox{Discrete}\{1,\ldots,K_{\mbox{max}}\}\\
\label{eq:priorP}
\bs p|K &\sim \mbox{Dirichlet}(\gamma_1,\ldots,\gamma_K)\\
\label{eq:priorTheta}
\bs \theta_{kj}|K &\sim \mbox{Beta}(\alpha,\beta),
\end{alignat}
independent for $k = 1,\ldots,K;j=1,\ldots,d$. The discrete distribution in Equation \eqref{eq:priorK} can be either a Uniform or a Poisson distribution with mean $\lambda = 1$ truncated on the set $\{1,\ldots,K_{\mbox{max}}\}$. Equations \eqref{eq:priorP} and \eqref{eq:priorTheta} correspond to typical prior distributions for the mixture weights and success probabilities, that furthermore enjoy conjugacy properties. Typically, we set $\gamma_1=\ldots=\gamma_K = \gamma > 0$ so that the prior assumptions do not impose any particular information that separates the mixture components between them, which is also a recommended practice in mixture modelling.

\begin{figure}[t]
\centering
\includegraphics[scale=0.35]{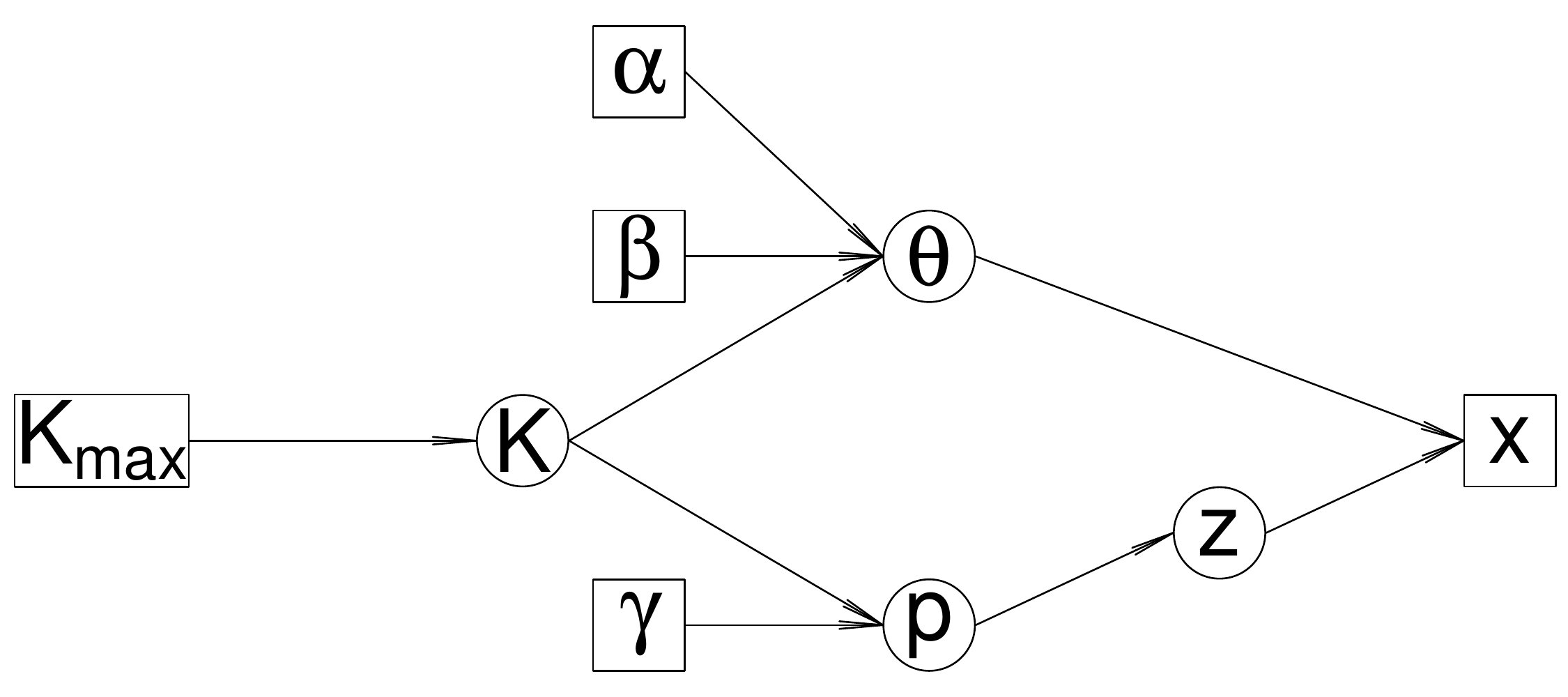}
\caption{Representation of the hierarchical model \eqref{eq:model} as a directed acyclic graph. Squares and circles denote observed/fixed and unknown variables, respectively.}
\label{fig:dag}
\end{figure}

According to Equations \eqref{eq:priorZ}, \eqref{eq:completeL}, \eqref{eq:priorK}, \eqref{eq:priorP} and \eqref{eq:priorTheta}, the joint probability density function of the model is
\begin{equation}
f\left(\bs x, K, \bs z, \bs p, \bs\theta\right) = f\left(\bs x|K,\bs z, \bs\theta\right)f\left(\bs z|K,\bs p\right)f\left(\bs p|K\right)f\left(\bs\theta|K\right)f\left(K\right),\label{eq:model}
\end{equation}
and its graphical representation is shown in Figure \ref{fig:dag}.

\section{Inference}\label{sec:inference}
\subsection{Allocation sampler}\label{sec:sampler}

Let $C_K = \frac{\Gamma(\sum_{k=1}^{K}\gamma_k)}{\prod_{k=1}^{K}\Gamma(\gamma_k)}
\left\{\frac{\Gamma(\alpha+\beta)}{\Gamma(\alpha)\Gamma(\beta)}\right\}^{Kd}$. From Equation \eqref{eq:model} the joint posterior distribution of $(K,\bs p,\bs\theta,\bs z)$ can be expressed as:
\begin{equation}
f(\bs\theta,\bs z, \bs p,K|\bs x) \propto C_K f(K)\prod_{k=1}^{K}\left\{p_k^{n_k + \gamma_k - 1}\prod_{j=1}^{d}\theta_{kj}^{\alpha+s_{kj}-1}(1-\theta_{kj})^{\beta+n_k-s_{kj} - 1}\right\}\mathbb{I}_{\mathcal A}(K,\bs p,\bs\theta,\bs z).
\label{eq:posterior}
\end{equation}
From the last expression it is straightforward to derive the full conditional distributions of the component specific parameters and latent allocation variables as follows:
\begin{alignat}{1}
\bs p|K,\bs z &\sim \mbox{Dirichlet}\left(\gamma_1 + n_1,\ldots,\gamma_K+n_K\right)\label{eq:conditionalP}\\
\theta_{kj}|K,\bs z,\bs x &\sim \mbox{Beta}\left(\alpha + s_{kj},\beta + n_k - s_{kj}\right)\label{eq:conditionalTheta}\\
\mathrm{P}\left(z_i = k|K,\bs x_i,\bs p, \bs \theta\right) &\propto p_k\prod_{j=1}^{d}\theta_{kj}^{x_{ij}}\left(1-\theta_{kj}\right)^{1-x_{ij}},\quad k=1,\ldots,K,\nonumber
\end{alignat}
independent for $i = 1,\ldots,n$; $k = 1,\ldots,K$; $j = 1,\ldots,d$.

A general framework for updating the number of mixture components ($K$) is given by trans-dimensional MCMC approaches, such as the reversible jump MCMC \citep{Green:95, Richardson:97, papRJ} or the Birth-Death MCMC \citep{stephens2000} methodologies. However, the conjugate prior assumptions used for the component specific parameters ($\bs p,\bs\theta$) allow us to use simpler techniques by integrating those parameters out from the model and perform collapsed sampling \citep{liu1994collapsed} on the space of $(K,\bs z)$. We use the allocation sampler \citep{Nobile2007} which introduced this sampling scheme for parametric families such that conjugate prior distributions exist and also applied it in the specific context of mixtures of normal distributions. This approach was recently followed by \citet{white2016bayesian} which also allowed for variable selection.  

Let $\mathcal A_0 = \Theta^{Kd}\times\mathcal P_{K-1}$ and $\mathcal A_1 = \mathcal K\times\{1,\ldots,K\}^n$. Integrating out $(\bs\theta,\bs p)$ from \eqref{eq:posterior} we obtain
\begin{alignat}{1}\nonumber
f\left(K,\bs z|\bs x\right) &= \int\limits_{\mathcal A_0}f(\bs z, \bs\theta,\bs p,K|\bs x)\mathrm{d}\bs\theta\mathrm{d}\bs p\\
&\propto  C_K f(K)\mathbb{I}_{\mathcal{A}_1}(K,\bs z)\int\limits_{\mathcal A_0}\prod_{k=1}^{K}p_k^{n_k + \gamma_k - 1}\prod_{j=1}^{d}\theta_{kj}^{\alpha+s_{kj}-1}\left(1-\theta_{kj}\right)^{\beta+n_k-s_{kj} - 1}\mathrm{d}\bs\theta\mathrm{d}\bs p \nonumber\\
&\propto C_K f(K)\frac{\prod_{k=1}^{K}\Gamma\left(n_k+\gamma_k\right)}{\Gamma\left(n+\sum_{k=1}^{K}\gamma_{k}\right)}
\prod_{k=1}^{K}\prod_{j=1}^{d}\frac{\Gamma\left(\alpha+s_{kj}\right)\Gamma\left(\beta+n_k-s_{kj}\right)}{\Gamma\left(\alpha+\beta+n_k\right)}\mathbb{I}_{\mathcal{A}_1}(K,\bs z). \label{eq:collapsedPosterior}
\end{alignat}
Let now  $\bs z_{[-i]} = \{z_1,\ldots,z_{i-1},z_{i+1},\ldots,z_n\}$ and also define the following quantities, for $i = 1,\ldots,n$: 
\begin{alignat*}{1}
n_k^{[i]} &= \sum_{h\neq i}\mathbb{I}(z_h=k), k = 1,\ldots,K\\
s_{kj}^{[i]} &= \sum_{h\neq i}\mathbb{I}(z_h=k)x_{hj}, k = 1,\ldots,K; j = 1,\ldots,d\\
A_1^{[i]}&=\{j=1,\ldots,d:x_{ij}=1\}\\
A_0^{[i]}&=\{j=1,\ldots,d:x_{ij}=0\}.
\end{alignat*}
From Equation \eqref{eq:collapsedPosterior}, the (collapsed) conditional posterior distribution of $z_i$ is 
\begin{equation}\label{eq:zfull}
\mathrm{P}\left(z_{i}=k|\bs z_{[-i]},K,\bs x\right)\propto\frac{n_k^{[i]}+\gamma_k}{\left(\alpha+\beta+n_k^{[i]}\right)^{d}}\prod_{j\in A_1^{[i]}}\left(\alpha+s_{kj}^{[i]}\right)\prod_{j\in A_0^{[i]}}\left(\beta+n_k-s_{kj}^{[i]}\right),
\end{equation}
$k = 1,\ldots,K$; $i = 1,\ldots,n$.

It is well known that draws from the conditional distributions in Equation \eqref{eq:zfull} exhibit strong serial correlation, slowing down the convergence of the MCMC sampler. The mixing can be improved by proposing simultaneous updates of blocks of $\bs z|K$, by incorporating proper Metropolis-Hastings moves on $\bs z|K$. Following \cite{Nobile2007}, we also propose jumps to configurations that massively update the allocation vector as follows:
\begin{enumerate}
\item \textbf{Move 1:} select two mixture components and propose a random reallocation of the assigned observations.
\item \textbf{Move 2:} select two mixture components and propose to move a randomly selected subset of observations from the 1st to the 2nd one.
\item \textbf{Move 3:} select two mixture components and propose a reallocation of the assigned observations according to the full conditional probabilities given the already processed ones.
\end{enumerate}
Each move is accepted according to the corresponding Metropolis-Hastings acceptance probability, see \citet{Nobile2007} for details.

The final step of the allocation sampler is to update the number of clusters ($K$).  According to \citet{Nobile2007}, this is achieved by performing a Metropolis-Hastings type move, namely a pair of absorption/ejection moves which decrease/increase $K$, respectively. Assume that the current state of chain is $\{K,\bs z\}$. The following pseudocode describes the Absorption/Ejection step:
\begin{enumerate}
\item Attemp ejection with probability $p^{e}_K$, where $p^{e}_K = 1/2$, $K = 2,\ldots,K_{\max}-1$, $p_1^{e}=1$ and $p_{K_{\max}}^{e}=0$. Otherwise, an absorption move is attempted.
\item Suppose that an ejection is attempted. The candidate state is $\{K',\bs z'\}$ with $K' = K+1$.
\begin{enumerate}
\item Propose reallocation of observations assigned to the ejecting component between itself and the ejected component according to the $\mbox{Beta}(\tilde\alpha,\tilde\alpha)$ distribution.
\item Accept the candidate state with probability $\min\{1,R\}$ where
\begin{equation}\label{eq:ejectionRatio}
R = R(\tilde\alpha) = \frac{f\left(K',\bs z'|\bs x\right)}{f\left(K,\bs z|\bs x\right)}\frac{\mathrm{P}(\{K',\bs z'\}\rightarrow\{K,\bs z\})}{\mathrm{P}\left(\{K,\bs z\}\rightarrow\{K',\bs z'\}\right)}
\end{equation}
\end{enumerate}
\item If an absorption is attempted:
\begin{enumerate}
\item all observations allocated to the absorbed component are reallocated to the absorbing component.
\item the candidate state is accepted with probability $\min\{1,1/R(\tilde\alpha)\}$.
\end{enumerate}
\end{enumerate}
The parameter $\tilde\alpha$ is chosen in a way that ensures that the probability of ejecting an empty component is sufficiently large. For full details the reader is referred to \citet{Nobile2007}.

The allocation sampler for mixtures of multivariate Bernoulli distributions is summarized in the following algorithm.

\begin{algorithm}[Allocation sampler for Bernoulli mixtures]
Given an initial state $\{K^{(0)},\bs z^{(0)}\}\in\mathcal A_1$ iterate the following steps for $t = 1,2,\ldots$
\begin{enumerate}
\item For $i = 1,\ldots,n$
\begin{enumerate}
\item Compute $n_k^{[i]} = \sum_{h\neq i}\mathbb I(z_{h}=k)$, $s_{kj}^{[i]} = \sum_{h\neq i}\mathbb I(z_{h}=k)x_{hj}$, $k = 1,\ldots,K^{(t)}$; $j = 1,\ldots,d$.
\item Update $z_i^{(t)}|\bs z_{[-i]},\cdots$ according to Equation \eqref{eq:zfull}.
\end{enumerate}
\item Propose Metropolis-Hastings moves $M_1$, $M_2$ and $M_3$ to update $\bs z^{(t)}$.
\item Propose an Absorption/Ejection move to update $\{K^{(t)},\bs z^{(t)}\}$.
\end{enumerate}
\end{algorithm}
Note in step \textit{1.(a)}: 
\begin{equation*}
z_h = 
\begin{cases}
z_{h}^{(t)}, & h < i\\
z_{h}^{(t-1)}, & h > i.
\end{cases}
\end{equation*}
Finally, we mention that after the last step of Algorithm 1 we can also simulate the component-specific parameters $\bs p$ and $\bs\theta$ from their full conditional posterior distributions given in \eqref{eq:conditionalP} and \eqref{eq:conditionalTheta}, respectively. Although this is not demanded in case that the user is only interested in inferring $K,\bs z|\bs x$, it will produce an (approximate) MCMC sample from the full posterior distribution of $K,\bs p, \bs\theta,\bs z|\bs x$. If the observed data contains missing entries an extra step is implemented in order to simulate the corresponding values. For this purpose we use the full conditional distribution derived from Equation \eqref{eq:XgivenZ}, taking only into account the subset of $\{1,\ldots,d\}$ that contains missing values for a given $i = 1,\ldots,n$.

\subsection{Label switching issue and identifiability}\label{sec:labelSwitching}

Label switching \citep{redner1984mixture} is a well known identifiability problem occurring in MCMC outputs of mixture models, arising from the symmetry of the likelihood with respect to permutations of components' labels. A set of sufficient conditions under a general framework of missing data models that lead to label switching and its consequences is given in \citet{Papastamoulis2013}. If an MCMC sample exhibits label switching, the standard practice of estimating the posterior means and other parametric functions by ergodic averages becomes meaningless. In order to deal with this identifiability problem we have considered two versions of ECR algorithm \citep{Papastamoulis:10, papastamoulis2014handling, rodriguez} as well as the KL algorithm \citep{stephens2000dealing}. These algorithms are quite efficient and in most cases exhibit almost identical results, but ECR  is significantly faster and computationally lightweight compared to KL. The implementation was performed in the {R} package \CRANpkg{label.switching} \citep{papastamoulis2016label}. 

Note here that in the case that $d = 1$, Equation \eqref{eq:mixture} collapses to a single Bernoulli distribution. Hence, there are two types of identifiability issues in mixture models: the first one is related to the fact that the model is identifiable only up to a permutation of the parameters (label switching). The second one is strict non-identifiability which relates to the fact that for a mixture of discrete distributions (such as the multivariate Bernoulli) totally different parameter values can correspond to the same distribution. We are not dealing with this second source of identifiability problems since it has been empirically demonstrated that estimation can still produce meaningful results in practice \citep{carreira2000practical}. In addition, \citet{allman2009identifiability} showed that finite mixtures of Bernoulli products are in fact generically identifiable despite their lack of strict identifiability.

\subsection{Metropolis-coupled MCMC sampler}\label{sec:mc3}

There are various strategies for improving MCMC sampling, see e.g.~chapter 6 in \citep{Gilks96}. In this study, the Metropolis-coupled MCMC ($\mbox{MC}^{3}$) \citep{geyer1991, geyer1995, Altekar12022004} strategy is adopted. An $\mbox{MC}^{3}$ sampler runs $m$ chains with different posterior distributions $f_i(\xi); i=1,\ldots,m$. The target posterior distribution corresponds to $i=1$, that is, $f_1(\xi) = f(\xi)$, while the rest of them are chosen in a way that the mixing is improved. This is typically achieved by considering ``heated'' versions of the original target, that is, $f_i(\xi) = f(\xi)^{h_i}$ where $h_1 = 1$ and $0<h_i < 1$ for $i = 2,\ldots,m$ represents the heat value of the chain. Note that when raising the posterior distribution to a power $0<h_i<1$ makes the modified posterior surface flatter, thus, easier to explore compared to $f(\xi)$. Only the chain that corresponds to the target posterior distribution is used for the posterior inference, however, after each iteration a proposal attempts to swap the states of two randomly chosen chains. This improves the mixing of the chain since it is possible that an accepted swap between the cold and a heated chain will make the former move to another mode. 

\begin{figure}[t]
\centering
\includegraphics[scale=0.4]{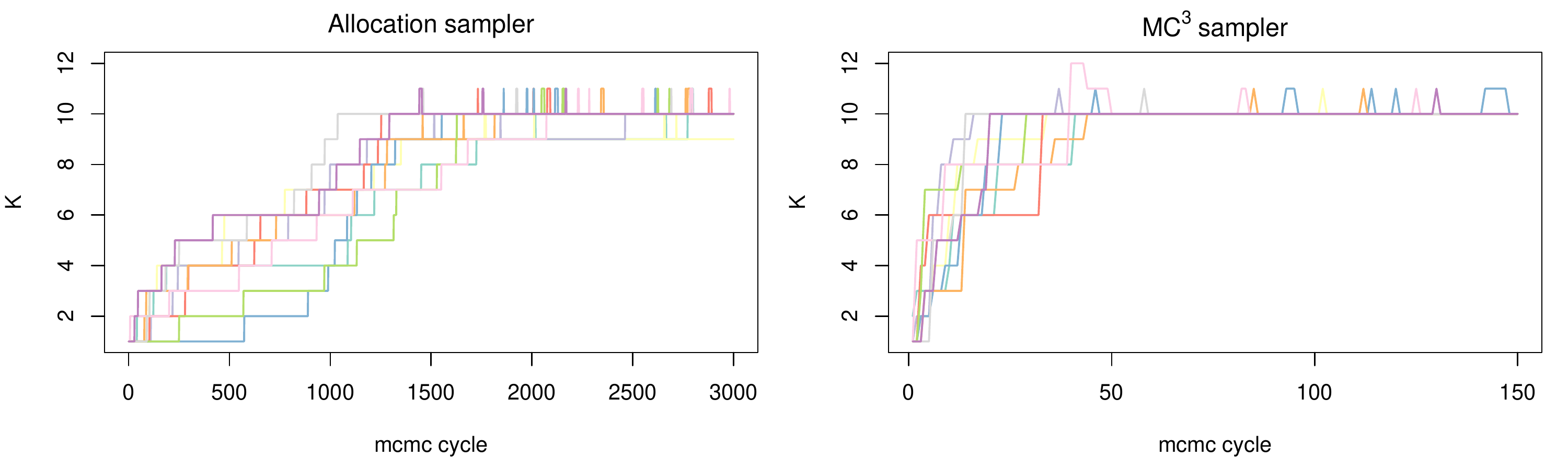}
\caption{Trace of sampled values of the number of clusters ($K$) for 10 different runs, using one of the synthetic datasets of Section \ref{sec:sim} with $K_{\mbox{true}} = 10$. Each run was initialized from $K = 1$ and every 10th iteration is displayed (labeled as MCMC cycle on the $x$ axis) until most chains start to explore $K = 10$. Left: standard allocation sampler , right: $\mbox{MC}^{3}$ sampler (using 4 heated chains).}
\label{fig:asVSmc3}
\end{figure}

Let $\xi^{(t)}_i$ denote the state of chain $i$ at iteration $t$ and that a swap between chains $i$ and $j$ is proposed. Note that in our setup $\xi = (K,\bs z)$ and $f$ is given in \eqref{eq:collapsedPosterior} (up to a normalizing constant). 
The proposed move is accepted with probability
$$\min\left\{1,\frac{f_i\left(\xi_j^{(t)}\right)f_j\left(\xi_i^{(t)}\right)}{f_i\left(\xi_i^{(t)}\right)f_j\left(\xi_j^{(t)}\right)}\right\}=
\min\left\{1,\frac{f\left(\xi_j^{(t)}\right)^{h_i}f\left(\xi_i^{(t)}\right)^{h_j}}{f\left(\xi_i^{(t)}\right)^{h_i}f\left(\xi_j^{(t)}\right)^{h_j}}\right\}.$$
Figure \ref{fig:asVSmc3} sketches the difference in convergence speed between the standard allocation sampler and an $\mbox{MC}^3$ sampler which are used to infer the same posterior distribution of the number of clusters. Although a single MCMC cycle of $\mbox{MC}^3$ is more expensive than a cycle of the allocation sampler, it is evident that the $\mbox{MC}^3$ sampler can recover the true number of clusters ($K_{\mbox{true}} = 10$) in a remarkably smaller number of iterations than the standard allocation sampler. In addition, the standard allocation sampler rarely switches between the symmetric modes of the posterior distribution, a fact which typically indicates poor mixing of MCMC samplers in mixture models \citep{Marin:05}. On the contrary, the $\mbox{MC}^3$ sampler produces a chain where label switching occurs in a rate proportional to the swap acceptance rate. 

In order to take full advantage of computing power in modern-day computers, our $\mbox{MC}^3$ sampler utilizes parallel computing in multiple cores. This is achieved by running each chain in parallel using the {\tt R} packages \CRANpkg{foreach} \citep{foreach} and \CRANpkg{doParallel} \citep{doparallel}. Every 10-th iteration a swap is proposed between a pair of chains.

\section{Using package \pkg{BayesBinMix}}\label{sec:cmFunction}

The main function of the \pkg{BayesBinMix} package is \code{coupledMetropolis}, with its arguments shown in Table \ref{tab:input}. This function takes as input a binary data array (possibly containing missing values) and runs the allocation sampler for a series of heated chains which run in parallel while swaps between pairs of chains are proposed. In the case that the most probable number of mixture components is larger than 1, the label switching algorithms are applied.

\begin{table}[t]
\begin{tabular}{lp{11cm}}
\toprule
Argument & Description\\ \midrule
\code{Kmax} & Maximum number of clusters (integer, at least equal to two).\\
\code{nChains} & Number of parallel (heated) chains. \\
\code{heats} &  \code{nChains}-dimensional vector specifying the temperature of each chain:  the 1st entry should always be equal to 1 and the rest of them lie on the set:$(0,1]$.\\
\code{binaryData} & The observed binary data (array). Missing values are allowed as long as the corresponding entries are denoted as \code{NA}.\\
\code{outPrefix} & The name of the produced output folder.   An error is thrown if the directory exists.\\
\code{ClusterPrior} & Character string specifying the prior distribution of the number of clusters. Available options: \code{'poisson'} or \code{'uniform'}. It defaults to the (truncated) Poisson distribution.\\
\code{m} & The number of MCMC cycles.  At the end of each cycle a swap between a pair
of heated chains is attempted. Each cycle consists of 10 iterations.\\
\code{alpha} & First shape parameter of the Beta prior distribution (strictly positive). Defaults to 1.\\
\code{beta} & Second shape parameter of the Beta prior distribution (strictly positive).   Defaults to 1.\\
\code{gamma}& \code{Kmax}-dimensional vector (positive) corresponding to the parameters of the Dirichlet prior of the mixture weights. Default value: \code{rep(1,Kmax)}.\\
\code{z.true} & An optional vector of cluster assignments considered as the ground-truth clustering of the observations. It is only used to obtain a final permutation of the labels (after the label switching algorithms) in order to maximise the similarity between the resulting estimates and the real cluster assignments. Useful for simulations.\\
\code{ejectionAlpha} & Probability of ejecting an empty component. Defaults to 0.2.\\
\code{burn} & Optional integer denoting the number of MCMC cycles that will be discarded as burn-in period.\\
\bottomrule
\end{tabular}
\caption{Arguments of the \code{coupledMetropolis} function.}
\label{tab:input}
\end{table}

As the function runs it prints some basic information on the screen such as the progress of the sampler as well as the acceptance rate of proposed swaps between chains. The output which is returned to the user mainly consists of \code{mcmc} objects, a class imported from the \CRANpkg{coda} package \citep{coda}. More specifically, the \code{coupledMetropolis} function returns the objects detailed in Table \ref{tab:output}. We note that this is just a subset of the full output of the sampler which consists of several additional quantities, such as the raw MCMC values corresponding to the whole set of generated values of $K$. Usually this information is not necessary to the average user, thus, it is saved to a separate set of files in the folder specified by \code{outPrefix}. 
\begin{table}[h]
\begin{tabular}{lp{9cm}}
\toprule
Object & Description\\ \midrule
       \code{K.mcmc} &
                object of class \code{mcmc} (see \pkg{coda} package) containing the simulated values (after burn-in) of the number of clusters for the cold chain.\\
             
        \code{parameters.ecr.mcmc} & 
                object of class \code{mcmc} containing the simulated values (after burn-in) of $\theta_{kj}$ (probability of success per cluster $k$ and feature $j$) 
and $\pi_k$ (weight of cluster $k$) for $k = 1,\ldots,K_{\mbox{map}}$; $j = 1,\ldots,d$, where $K_{\mbox{map}}$ denotes the most probable number of clusters. The output is reordered according to \code{ECR} algorithm.\\
        
        \code{allocations.ecr.mcmc} &
                object of class \code{mcmc} containing the simulated values (after burn-in) of $z_{i}$ (allocation variables) for $i = 1,\ldots,n$, given $K = K_{\mbox{map}}$. The output is reordered according to \code{ECR} algorithm.\\
        
        \code{classificationProbabilities.ecr} &
                data frame of the reordered classification probabilities per observation after reordering the most probable number of clusters with the \code{ECR} algorithm.\\
        
        \code{clusterMembershipPerMethod}&
                data frame of the most probable allocation of each observation after reordering the MCMC sample which corresponds to the most probable number of clusters according to \code{ECR}, \code{STEPHENS} and \code{ECR-ITERATIVE-1} methods.\\
        
        \code{K.allChains}&
                \code{m}$\times$\code{nChains} matrix containing the simulated values of the number of clusters ($K$) per chain.\\
                \code{chainInfo} & Number of parallel chains, cycles, burn-in period and acceptance rate of swap moves.\\
\bottomrule
\end{tabular}
\caption{Output returned to the user of the \code{coupledMetropolis} function.}
\label{tab:output}
\end{table}

\section{Examples}
In this section the usage of \pkg{BayesBinMix} package is described and various benchmarks are presented. At first we demonstrate a typical implementation on a single simulated dataset and inspect the simulated parameter values and estimates. Then we perform an extensive study on the number of estimated clusters and compare our findings to the \CRANpkg{FlexMix} package \citep{flexmix1,flexmix2,flexmix3}. An application to a real dataset is provided next.

\subsection{Simulation study}\label{sec:sim}

At first, a single simulated dataset is used in order to give a brief overview of the implementation. We simulated $n = 200$ observations from the multivariate Bernoulli mixture model \eqref{eq:mixture}. The true number of clusters is set to $K = 6$ and the dimensionality of the multivariate distribution is equal to $d = 100$. The mixture weights are drawn from a Dirichlet $\mathcal D(1,1,1,1,1,1)$ distribution resulting in $(50, 46, 30, 36, 12, 26)$ generated observations from each cluster. For each cluster, true values for the probability of success were generated from a Uniform distribution, that is, $\theta_{kj}\sim \mathcal U(0,1)$, independently for $k = 1,\ldots,K$; $j = 1,\ldots,d$. Furthermore, we introduce some missing values to the generated data: each row is allowed to contain missing values with probability $0.2$: for such a row the total number of missing entries is drawn from the binomial distribution $B(100,0.3)$. Finally, the observed data is saved to the $200\times 100$ array \code{x} which contains a total of 1038 missing values corresponding to 34 rows. 

We will run 4 parallel chains with the following temperatures: $(1,0.8,0.6,0.4)$. Observe that the first chain should correspond to the actual posterior distribution, so its temperature equals to 1. Now apply the \code{coupledMetropolis} function as follows.

\begin{example}
> library('BayesBinMix')
> nChains <- 4
> heats <- seq(1,0.4,length = nChains)

# using the truncated Poisson prior distribution on the number of clusters
> cm1 < - coupledMetropolis(Kmax = 20, nChains = nChains, heats =  heats, 
                binaryData = x, outPrefix = 'bbm-poisson', ClusterPrior = 'poisson', 
                m = 1100, z.true = z.true, burn = 100)

# using the uniform prior distribution on the number of clusters
> cm2 <- coupledMetropolis(Kmax = 20, nChains = nChains, heats =  heats, 
                binaryData = x, outPrefix = 'bbm-uniform', ClusterPrior = 'poisson', 
                m = 1100, z.true = z.true, burn = 100)
\end{example}

Note that we have called the function twice using either the truncated Poisson or the Uniform prior on the set $\{1,\ldots,20\}$. The total number of MCMC cycles corresponds to $m = 1100$ and the first $100$ cycles will be discarded as burn-in period. Recall that each cycle contains 10 usual MCMC iterations, so this is equivalent to keeping every 10th iteration of a chain with $11000$ iterations. Since we are interested to compare against the true values used to generate the data, we also supply \code{z.true} which contains the true allocation of each observation. It is only used for making the inferred clusters agree to the labelling of the true values and it has no impact on the MCMC or label switching algorithms.

\paragraph{Printing, summarizing and plotting the output:} In this section we  illustrate summaries of the basic output returned to the user, using only the run which corresponds to the Poisson prior distribution (\code{cm1}). The \code{print} method of the package returns a basic summary of the fitted model:

\begin{example}
> print(cm1)

* Run information: 
      Number of parallel heated chains: 4 
      Swap acceptance rate: 63.5% 
      Total number of iterations: 11000 
      Burn-in period: 1000 
      Thinning: 10. 

* Estimated posterior distribution of the number of clusters: 

    6     7     8 
0.971 0.026 0.003 

* Most probable model: K = 6 with P(K = 6|data) = 0.971 

* Estimated number of observations per cluster conditionally on K = 6 (3 label switching
  algorithms): 
  STEPHENS ECR ECR.ITERATIVE.1
1       50  50              50
2       46  46              46
3       30  30              30
4       36  36              36
5       12  12              12
6       26  26              26

* Posterior mean of probability of success per feature and cluster (ECR algorithm): 
         cluster_1 cluster_2 cluster_3 cluster_4  cluster_5 cluster_6
theta_1 0.33364058 0.8465393 0.7023264 0.3340989 0.08364937 0.8933767
theta_2 0.71919239 0.6653526 0.3227822 0.3982836 0.22369486 0.5936094
theta_3 0.49869339 0.2285653 0.3605507 0.3570447 0.07206039 0.1883581
theta_4 0.22360156 0.9148123 0.3359406 0.7889224 0.15476900 0.5924109
theta_5 0.01867034 0.8296381 0.8107050 0.1121773 0.78051586 0.1442368
   <+ 95 more rows> 
\end{example}

Next we present summaries of the marginal posterior distributions of the (reordered) MCMC sample of parameters conditionally on the selected number of clusters. The reordered MCMC sample of $\theta_{kj}$ and $p_k$ (after burn-in) is returned to the \code{mcmc} object \code{parameters.ecr.mcmc}. Hence we can use the \code{summary} method of the \pkg{coda} package, which prints empirical means, standard deviations, as well the quantiles for each variable. This is done with the following command. 

\begin{example}
> summary(cm1$parameters.ecr.mcmc)

1. Empirical mean and standard deviation for each variable,
   plus standard error of the mean:

               Mean      SD  Naive SE Time-series SE
theta.1.1   0.33364 0.06504 0.0020874      0.0020874
...         ...     ...     ...            ...
theta.6.1   0.89338 0.05552 0.0017816      0.0017816
   <+ 99 blocks of 6 rows>
p.1         0.24663 0.02869 0.0009208      0.0009208
...         ...     ...     ...            ...
p.6         0.13270 0.02276 0.0007304      0.0007304

2. Quantiles for each variable:

                 2.5%      25%     50%     75%   97.5%
theta.1.1   0.2115064 0.289971 0.32896 0.37502 0.46542
...         ...       ...      ...     ...     ...
theta.6.1   0.7676282 0.859599 0.90175 0.93405 0.97556
   <+ 99 blocks of 6 rows>
p.1         0.1950401 0.225938 0.24436 0.26611 0.31012
...         ...       ...      ...     ...     ...
p.6         0.0905194 0.117203 0.13206 0.14795 0.17993
\end{example}

\begin{figure}[t]
\begin{center}
\includegraphics[scale=0.45]{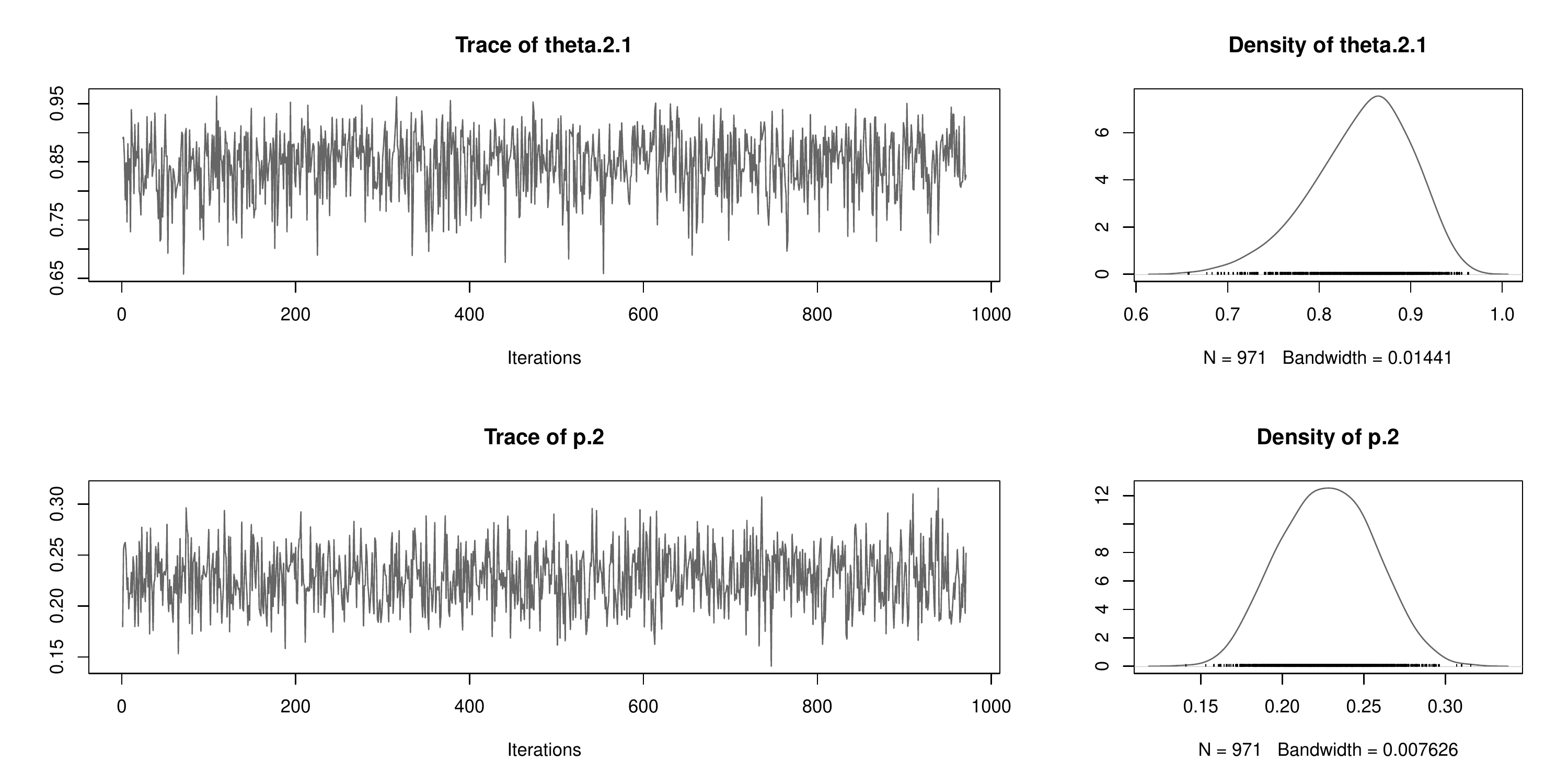}
\end{center}\caption{MCMC trace and density estimate for the reordered values of $\theta_{kj}$ and $p_k$ for cluster $k = 2$ and feature $j=1$, conditionally on the selected number of clusters ($K=6$).}
\label{fig:plotMethod}
\end{figure}

The user can also visualize the output with a trace of the sampled values and a density estimate for each variable in the chain using the \code{plot} method of the \pkg{coda} package. For illustration, the following example plots the trace and histogram for $\theta_{kj}$ and $p_k$ for cluster $k = 2$ and feature $j=1$. The produced plot is shown in Figure \ref{fig:plotMethod}.

\begin{figure}[p]
\begin{tabular}{cc}
\includegraphics[scale=0.33]{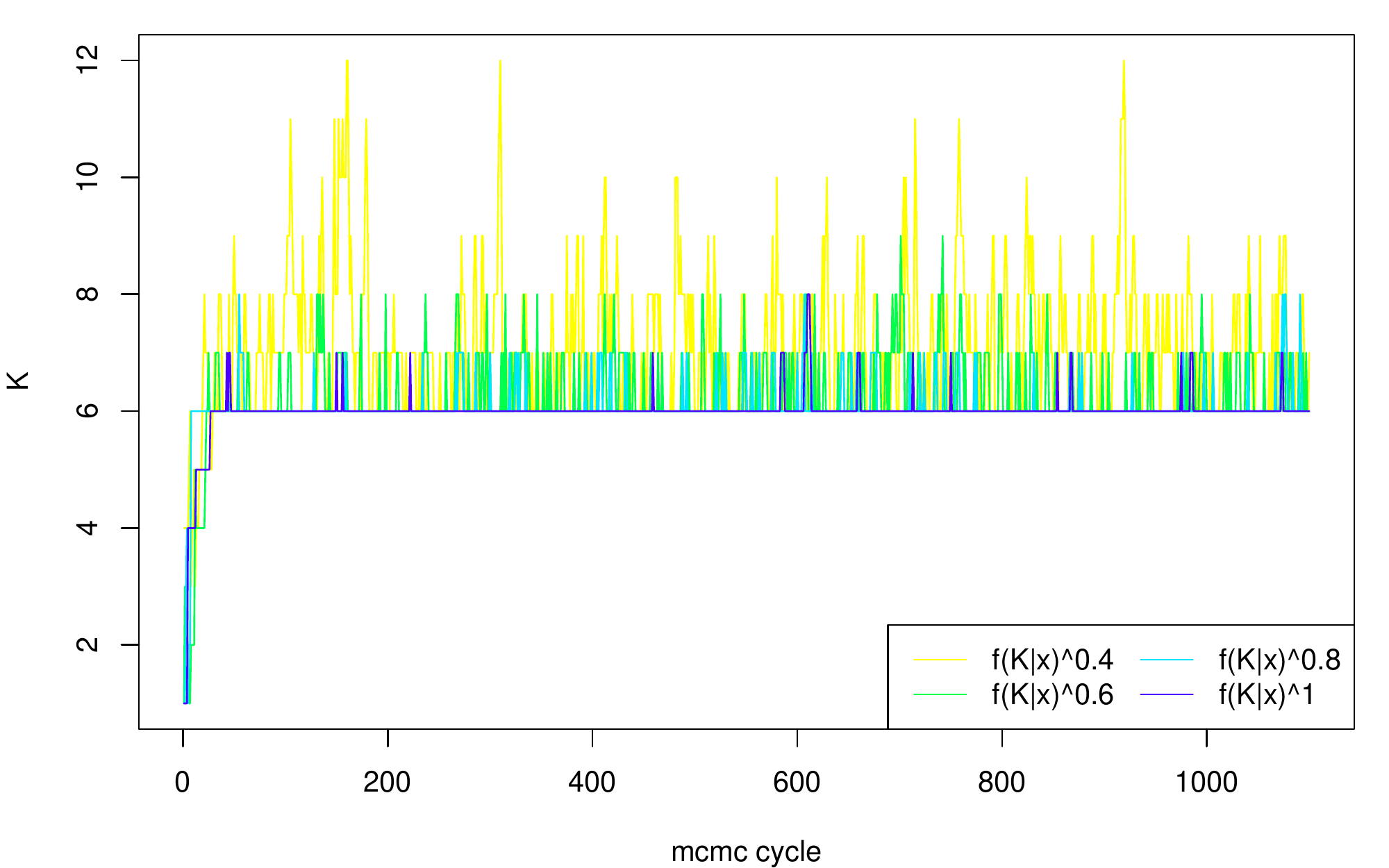}&
\includegraphics[scale=0.33]{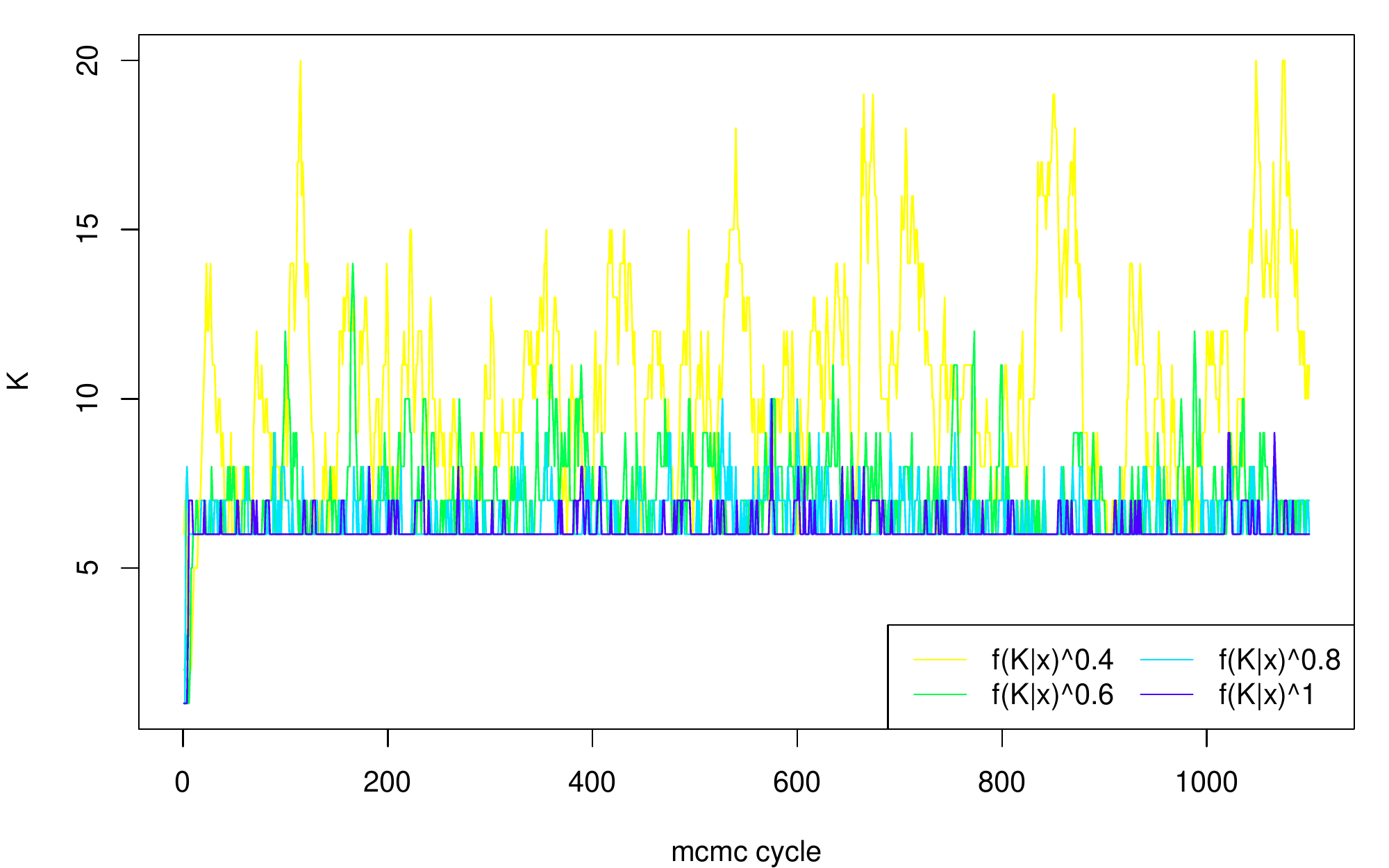}\\
(a) & (b)\\
\includegraphics[scale=0.33]{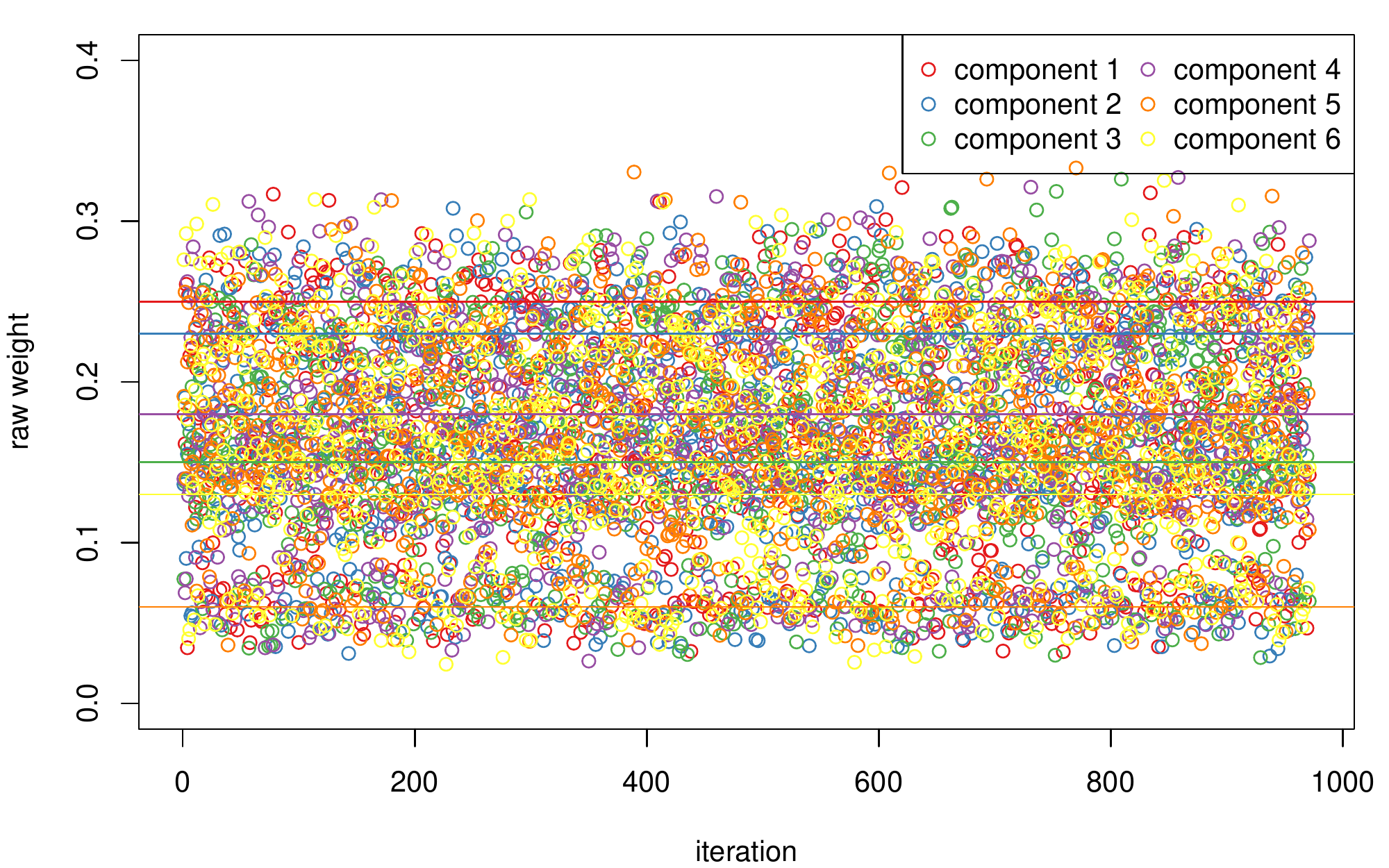}&
\includegraphics[scale=0.33]{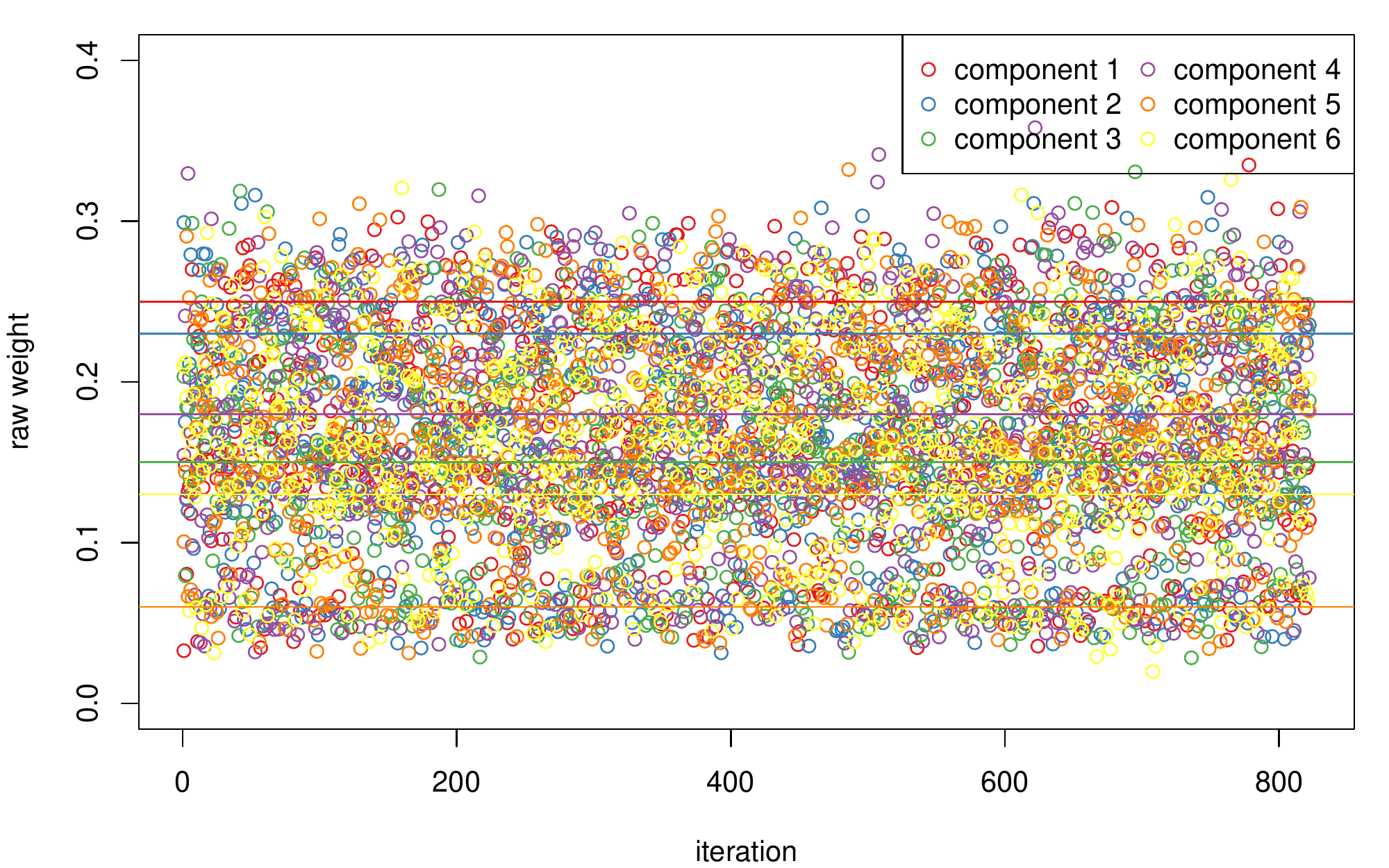}\\
(c) & (d)\\
\includegraphics[scale=0.33]{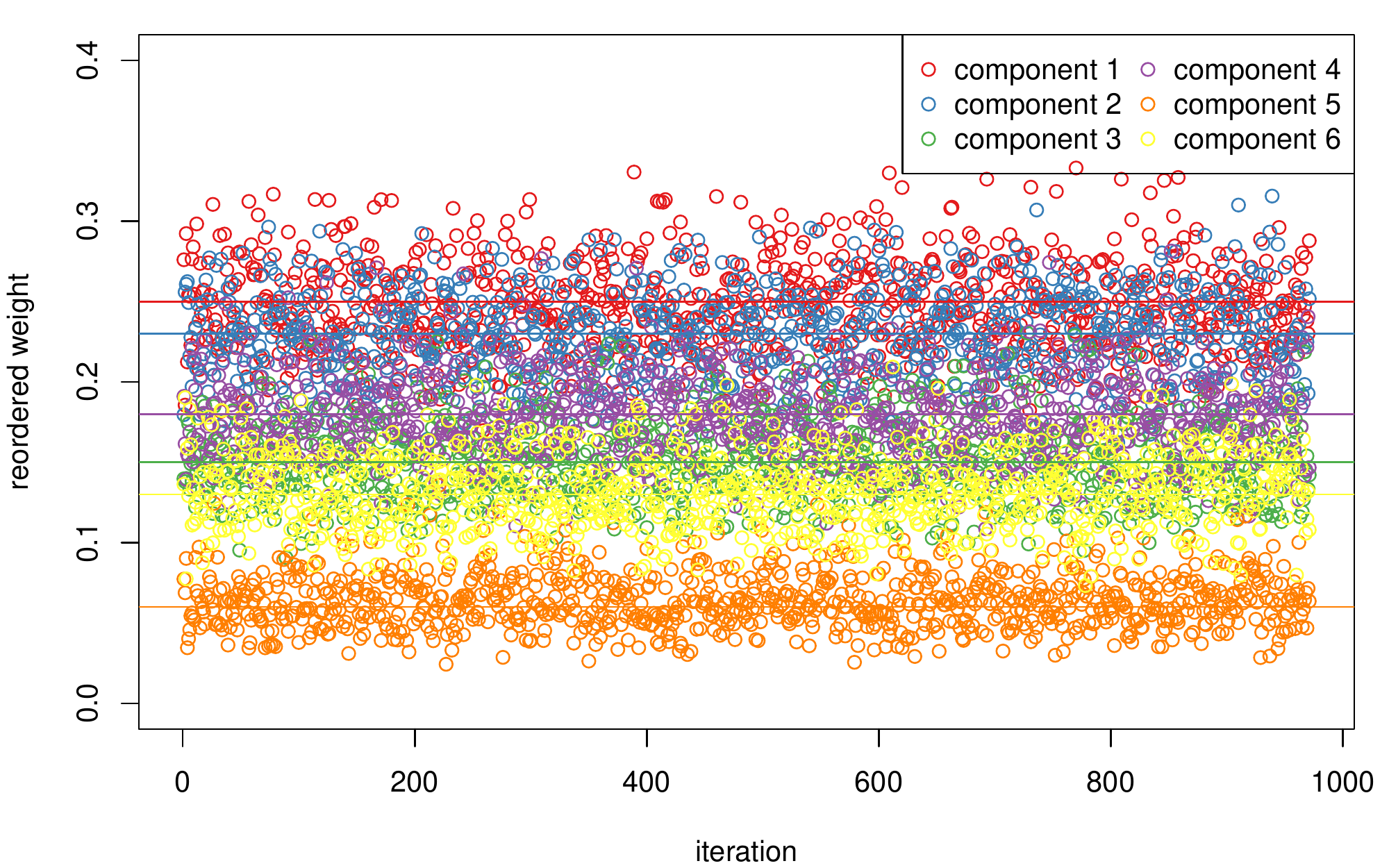}&
\includegraphics[scale=0.33]{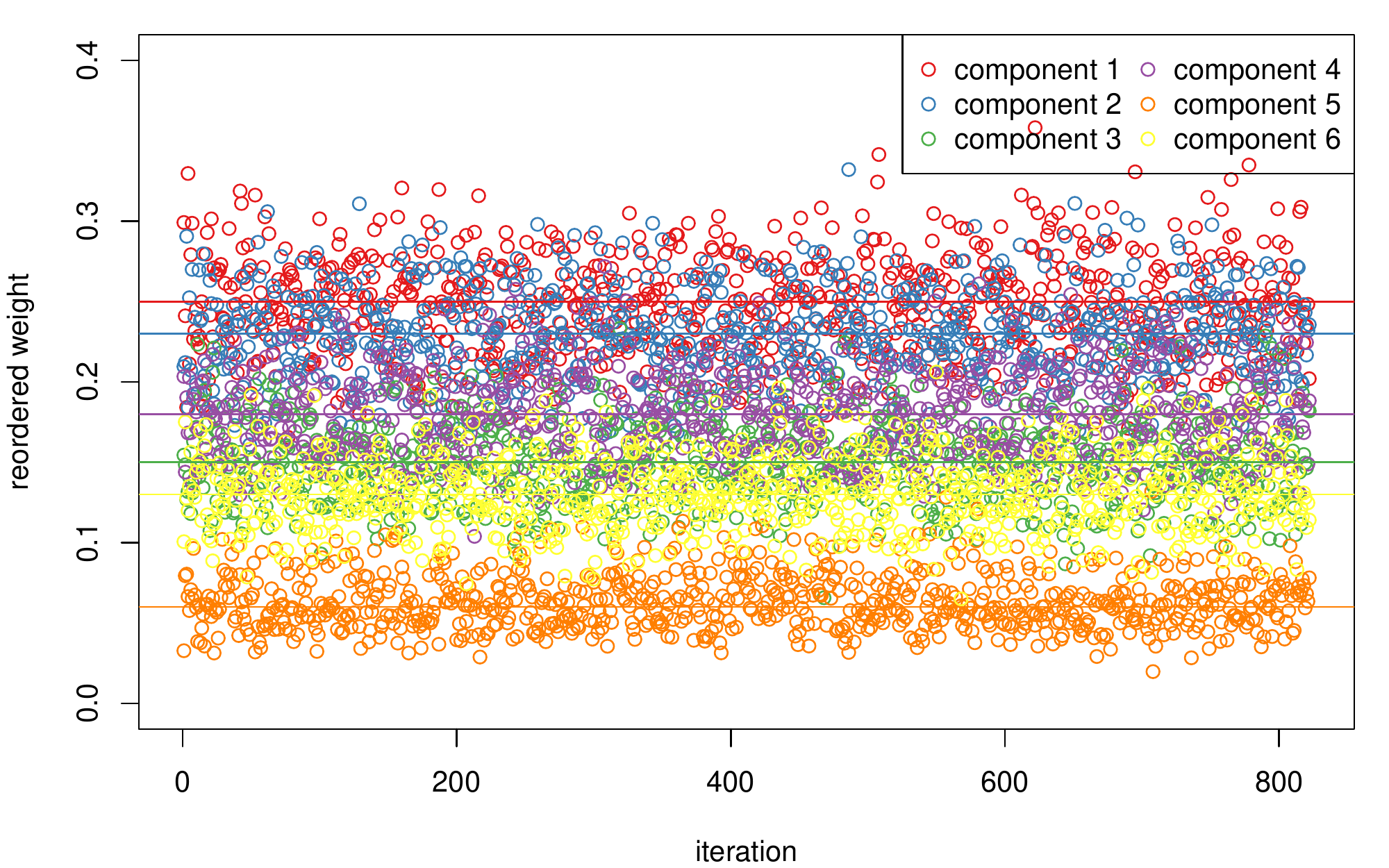}\\
(e) & (f)\\
\includegraphics[scale=0.33]{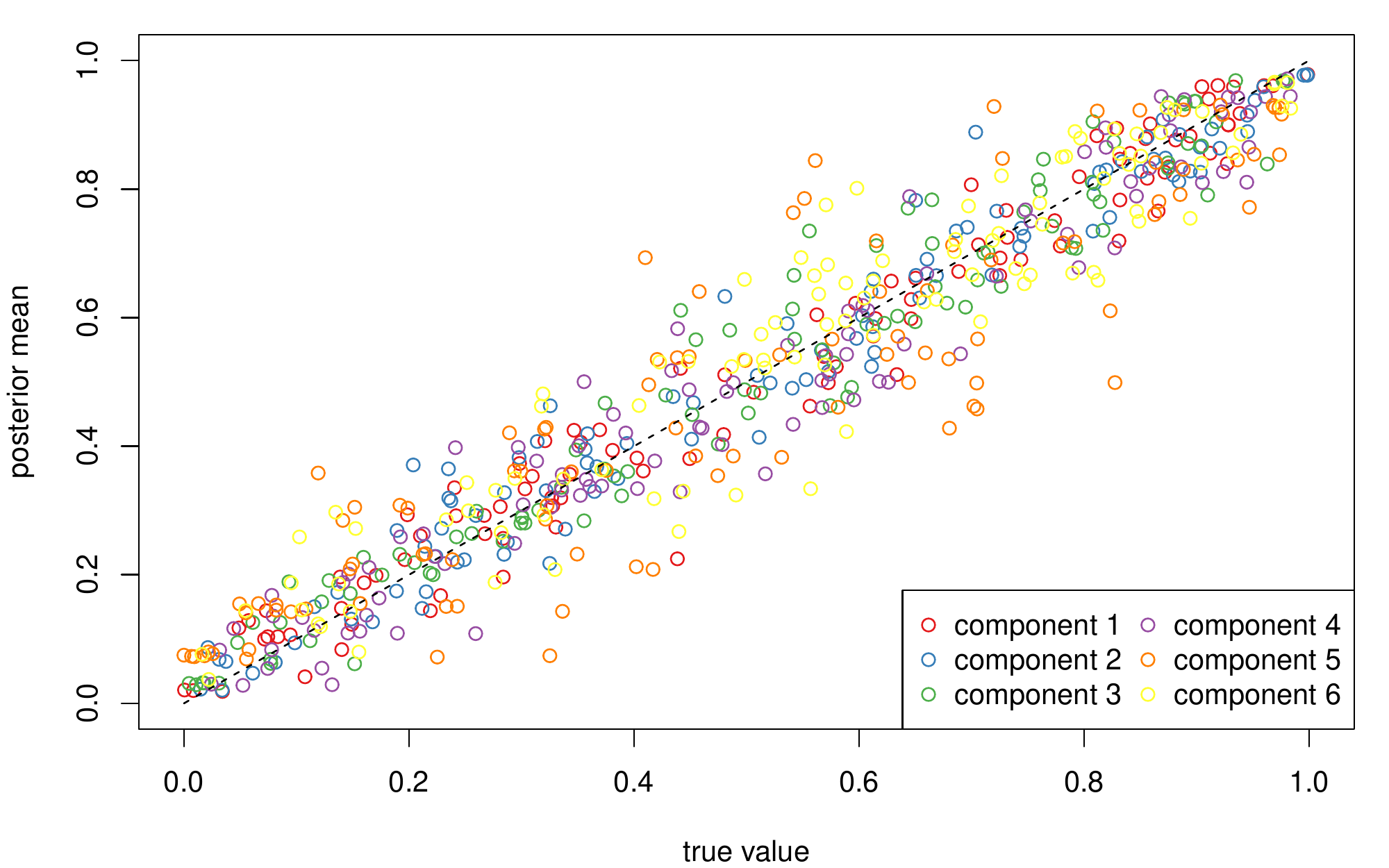}&
\includegraphics[scale=0.33]{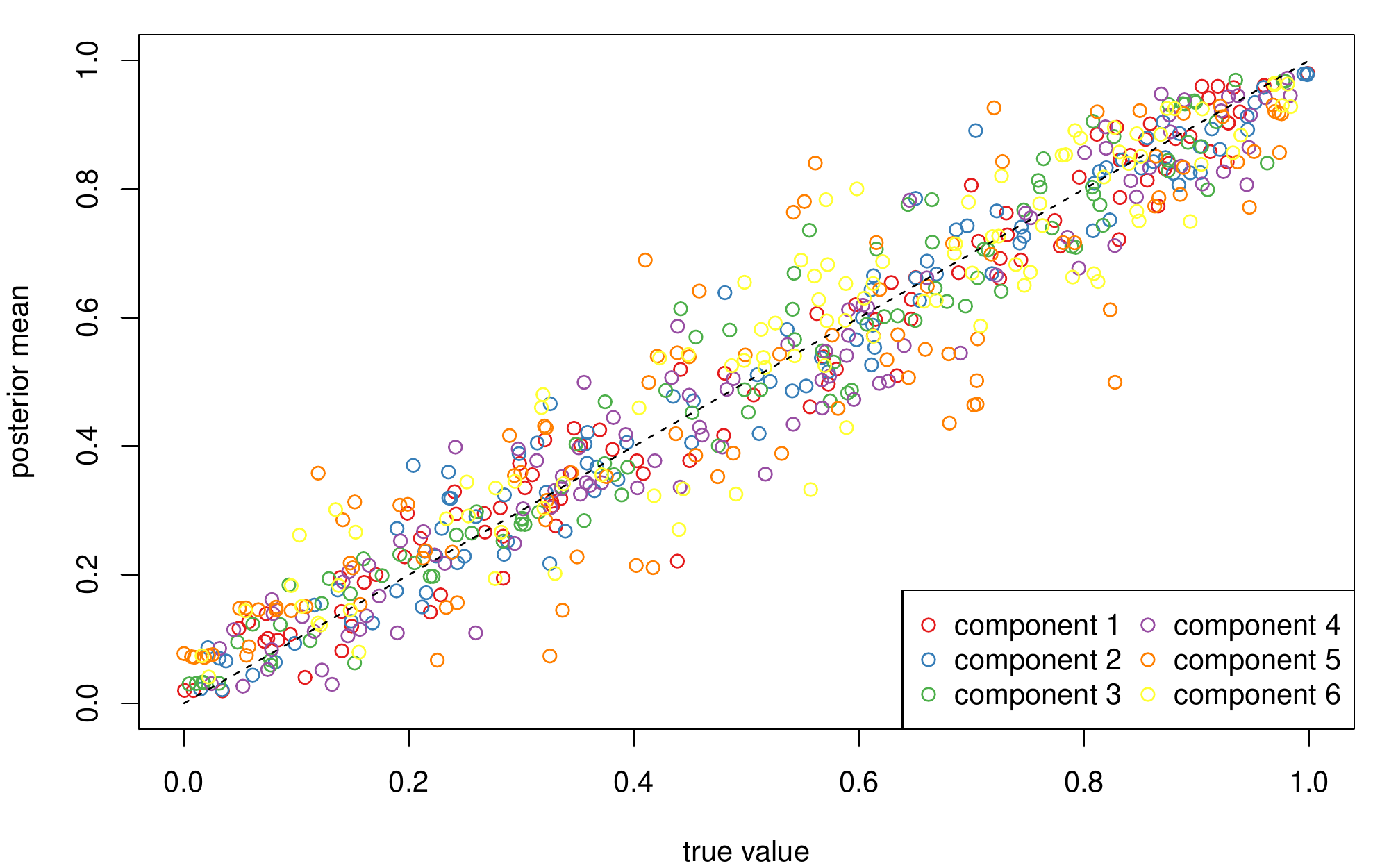}\\
(g) & (h)
\end{tabular}
\caption{Results for simulated data with $K_{\mbox{true}} = 6$ using the Poisson (left) or Uniform (right) prior distribution on the number of clusters ($K$). (a) and (b): generated values of $K$ per heated chain. (c) and (d): raw output of $p_1,\ldots,p_K$ conditionally on $K = 6$. (e) and (f): reordered sample according to ECR algorithm. Horizontal lines indicate true values of relative number of observations per cluster. (g) and (h): posterior mean estimates of Bernoulli parameters per cluster versus true values.}
\label{fig:bbm1}
\end{figure}

\begin{example}
mat <- matrix(c(1:4), byrow = TRUE, ncol = 2)
layout(mat, widths = rep(c(2,1), 2), heights = rep(1,4))
mcmcSubset <- cm1$parameters.ecr.mcmc[ , c("theta.2.1", "p.2")]
plot(mcmcSubset, auto.layout = FALSE, ask = FALSE, col = "gray40")
\end{example}
The reader is also referred to the \pkg{coda} package which provides various other functions for calculating and plotting MCMC diagnostics. 
\paragraph{Further inspection of the output:} Figures \ref{fig:bbm1}.(a) and \ref{fig:bbm1}.(b) illustrate the sampled values of $K$ per chain according to the Poisson and Uniform prior distribution, respectively. This information is returned to the user as an \code{m}$\times$\code{nChains} array named \code{K.allChains}. The actual posterior distribution corresponds to the blue line. Note that as the temperature increases the posterior distribution of $K$ has larger variability. In both cases, the most probable state corresponds to $K = 6$ clusters, that is, the true value. 

Next we inspect the MCMC output conditionally on the event that the number of clusters equals $6$ and compare to the true parameter values. At first, we can inspect the raw MCMC output, which is not identifiable due to the label switching problem. Thus, this information is not directly returned to the user, however it is saved to the file \file{rawMCMC.mapK.6.txt} in the output directory specified by the \code{output} argument. For illustration we plot the raw values of mixture weights. As shown in Figures \ref{fig:bbm1}.(c) and \ref{fig:bbm1}.(d), the sample is mixing very well to the symmetric posterior areas, since in every iteration labels are changing. The corresponding reordered values (according to the ECR algorithm) are returned to the user as an \code{mcmc} object named \code{parameters.ecr.mcmc}, shown in Figures \ref{fig:bbm1}.(e) and \ref{fig:bbm1}.(f). Note that the high posterior density areas are quite close to the true values of relative frequencies of generated observations per cluster (indicated by horizontal lines). Finally, Figures \ref{fig:bbm1}.(g) and \ref{fig:bbm1}.(h) display the posterior mean estimates (arising from the reordered MCMC sample) versus the true values of $\theta_{kj}$, $k = 1,\ldots,6$; $j = 1,\ldots,100$.

\begin{figure}[t]
\begin{tabular}{cc}
\includegraphics[scale=0.37]{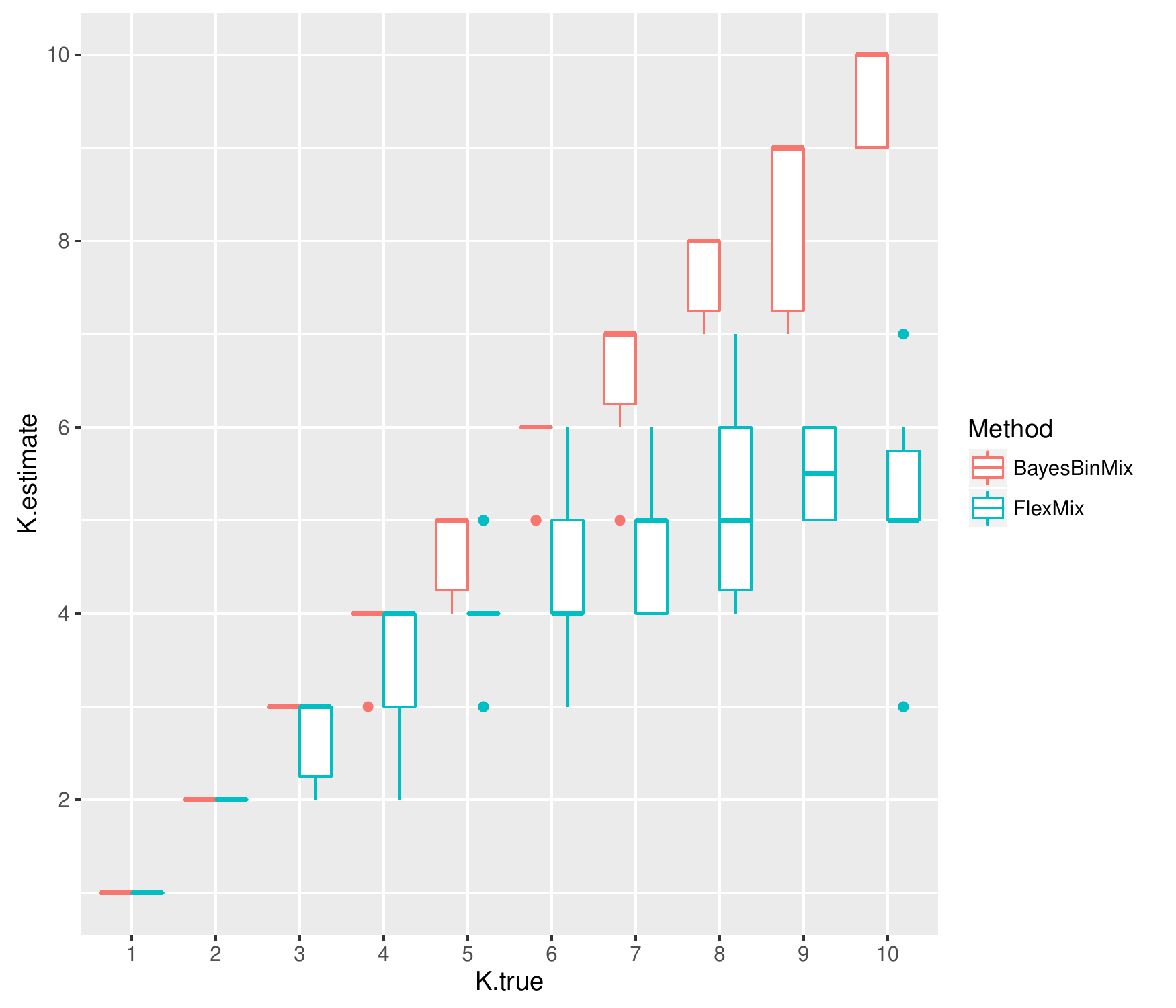}&
\includegraphics[scale=0.37]{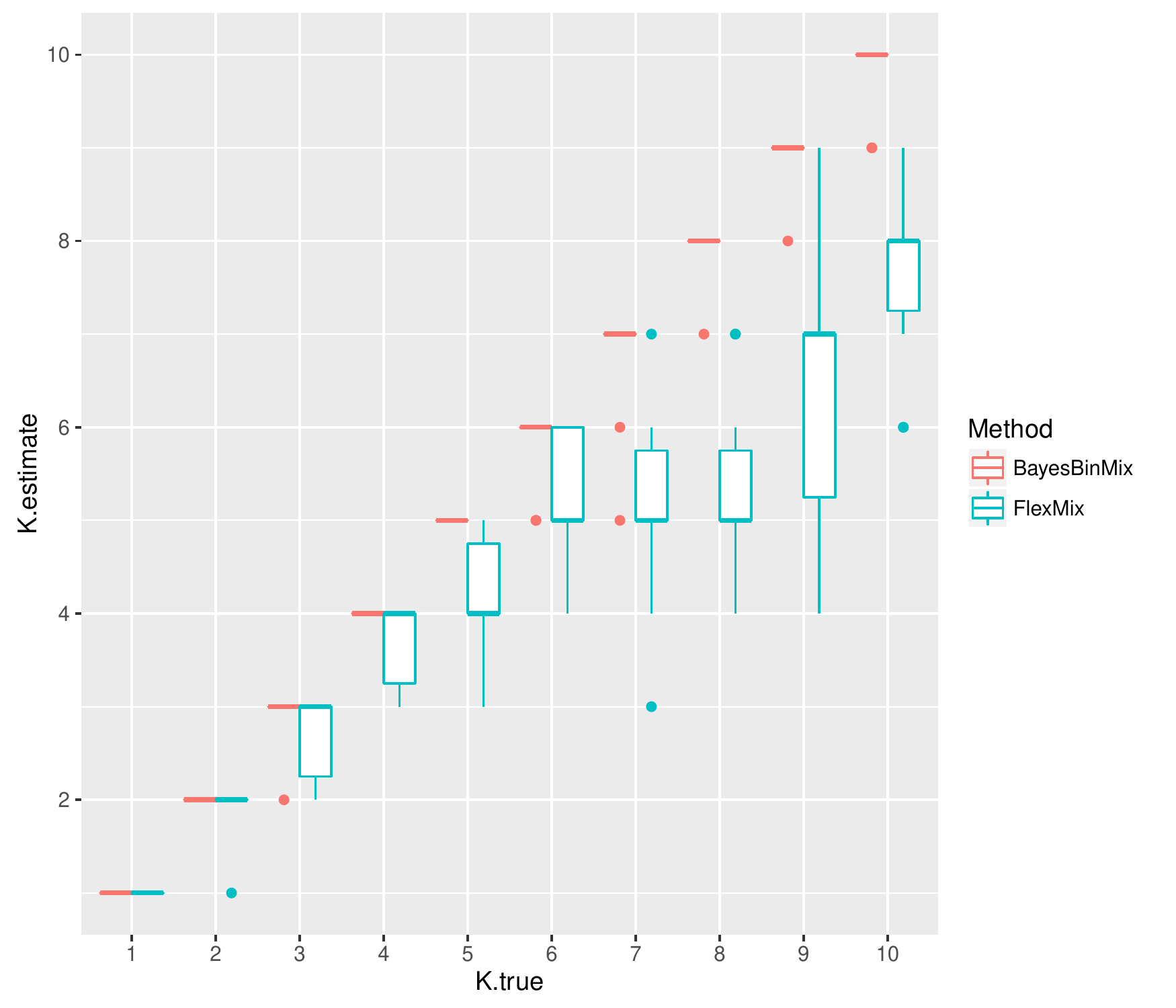}\\
(a) $n = 200$ & (b) $n = 300$\\
\includegraphics[scale=0.37]{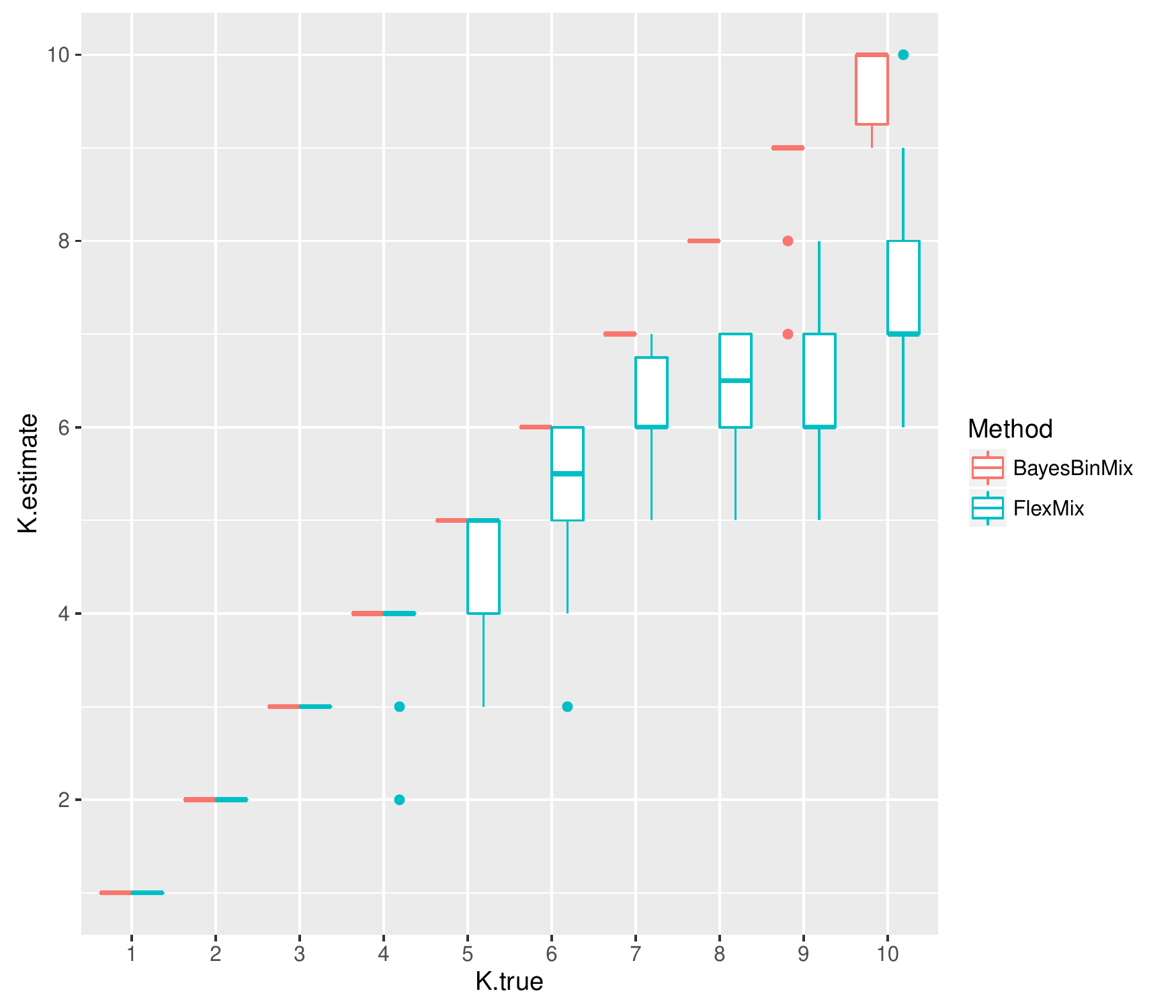}&
\includegraphics[scale=0.37]{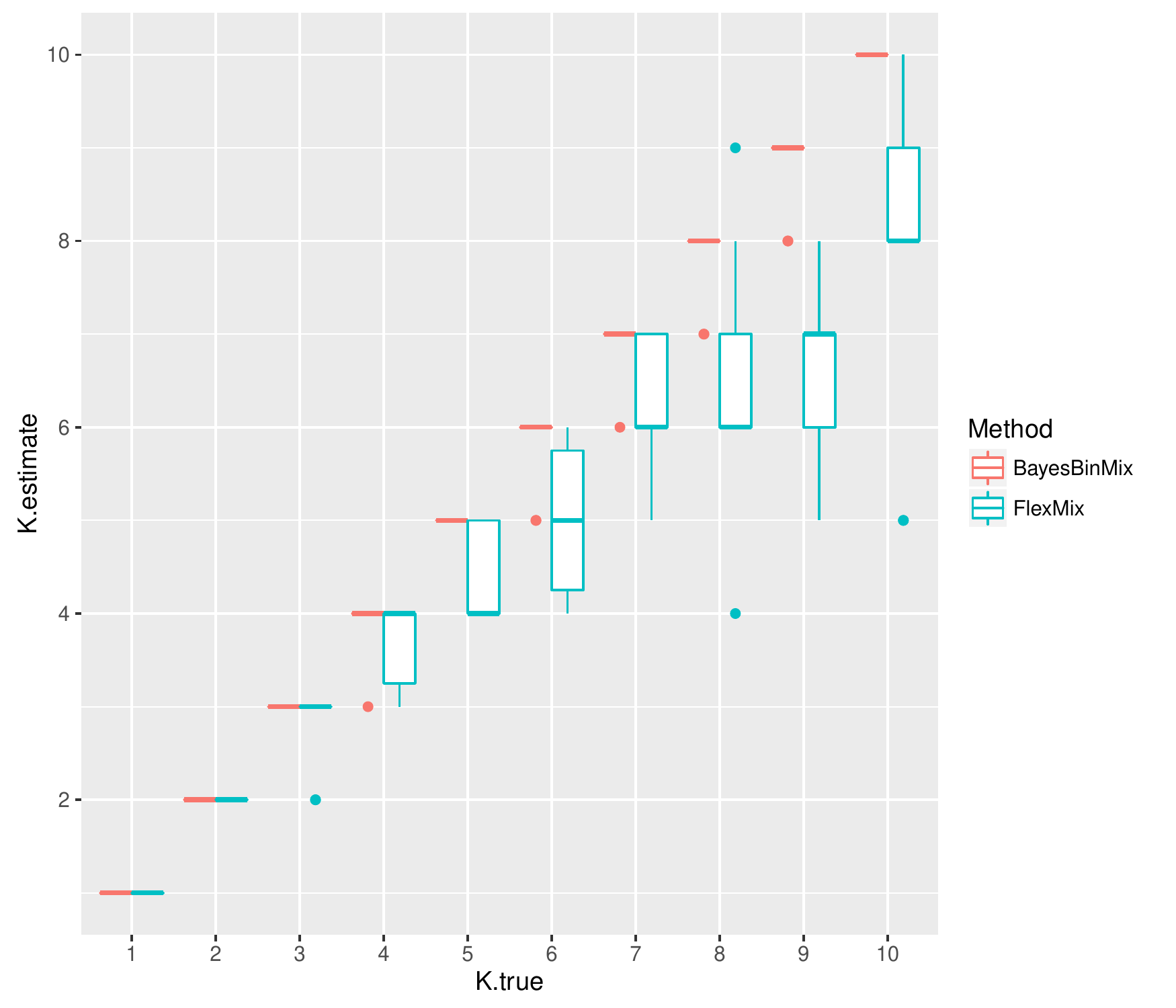}\\
(c) $n = 400$ & (d) $n = 500$
\end{tabular}
\caption{Model selection comparison between \pkg{BayesBinMix} and \pkg{FlexMix}. The x axis corresponds to the true number of clusters and the y axis to the estimated value. Each boxplot corresponds to 10 simulated datasets from a mixture of Bernoulli distributions.}
\label{fig:bbm_flexmix}
\end{figure}

\paragraph{Model selection study:} Next we are dealing with model selection issues, that is, selecting the appropriate number of clusters. For this reason we compare \pkg{BayesBinMix} with the EM-algorithm implementation provided in \pkg{FlexMix}. Under a frequentist framework, the selection of the number of mixture components is feasible using penalized likelihood criteria, such as the BIC \citep{Schwarz:78} or ICL \citep{Biernacki:2000}, after fitting a mixture model for each possible value of $K$. We used the ICL criterion since it has been shown to be more robust than BIC, see e.g.~\citet{papastamoulis2016estimation}. We considered that the true number of clusters ranges in the set $\{1,2,\ldots,10\}$ and for each case we simulated 10 datasets using the same data generation procedure as previously but without introducing any missing values due to the fact that \pkg{FlexMix} does not handle missing data. The number of observations varies in the set $n\in\{200,300,400,500\}$. For each simulated data the general call is the following.
\begin{example}
> library('BayesBinMix')
> library('flexmix')
> nChains <- 8
> heats <- seq(1,0.4,length = nChains)
> cm <- coupledMetropolis(Kmax = 20, nChains = nChains, heats =  heats, binaryData = x,
                outPrefix = 'sampler', ClusterPrior = 'poisson', m = 330, burn = 30)
# now run flexmix for binary data clustering
> ex <- initFlexmix(x ~ 1, k = 1:20, model = FLXMCmvbinary(), 
        control = list(minprior = 0), nrep = 10)
\end{example}

Note that for both algorithms the number of clusters varies in the set $\{1,\ldots,20\}$. Eight heated chains are considered for the MCMC scheme, while each run of the EM algorithm is initialised using \code{nrep = 10} different starting points in \pkg{FlexMix}. Here we used a total of only \code{m = 330} MCMC cycles in order to show that reliable estimates can be obtained using small number of iterations. Figure \ref{fig:bbm_flexmix} displays the most probable number of mixture components estimated by \pkg{BayesBinMix} and the selected number of clusters using \pkg{FlexMix}, for each possible value of the true number of clusters used to simulate the data. Observe that when the number of clusters is less than 5 both methods are able to estimate the true number of mixture components. However, \pkg{FlexMix} tends to underestimate the number of clusters when $K \geqslant 5$, while \pkg{BayesBinMix} is able to recover the true value in most cases.

\subsection{Real data}\label{sec:realdata}

We consider the zoo database available at the UC Irvine Machine Learning Repository \citep{Lichman:2013}. The database contains 101 animals, each of which has 15 boolean attributes and 1 discrete attribute (\code{legs}). The partition of animals into a total of 7 classes (mammal, bird, reptile, fish, amphibian, insect and invertebrate)  can be considered as the ground-truth clustering of the data, provided in the vector \code{z.ground\_truth}. Following \citet{li2005general}, the discrete variable \code{legs} is transformed into six  binary features, which correspond  to  0, 2, 4, 5, 6 and 8 legs, respectively. Also we eliminate one of the two entries corresponding to \code{frog}, as suggested by \citet{li2005general}. In total we consider an $100\times 21$ binary array \code{x} as the input data. 

Recall that the Bernoulli mixture in Equation \eqref{eq:mixture} assumes that each cluster consists of a product of independent Bernoulli distributions. Here this assumption is not valid due to the fact that the six new binary variables arising from \code{legs} are not independent: they should sum at 1. Nevertheless, it is interesting to see how our method performs in cases that the data is not generated by the assumed model. 

\begin{figure}[t]
\begin{tabular}{cc}
\includegraphics[scale=0.35]{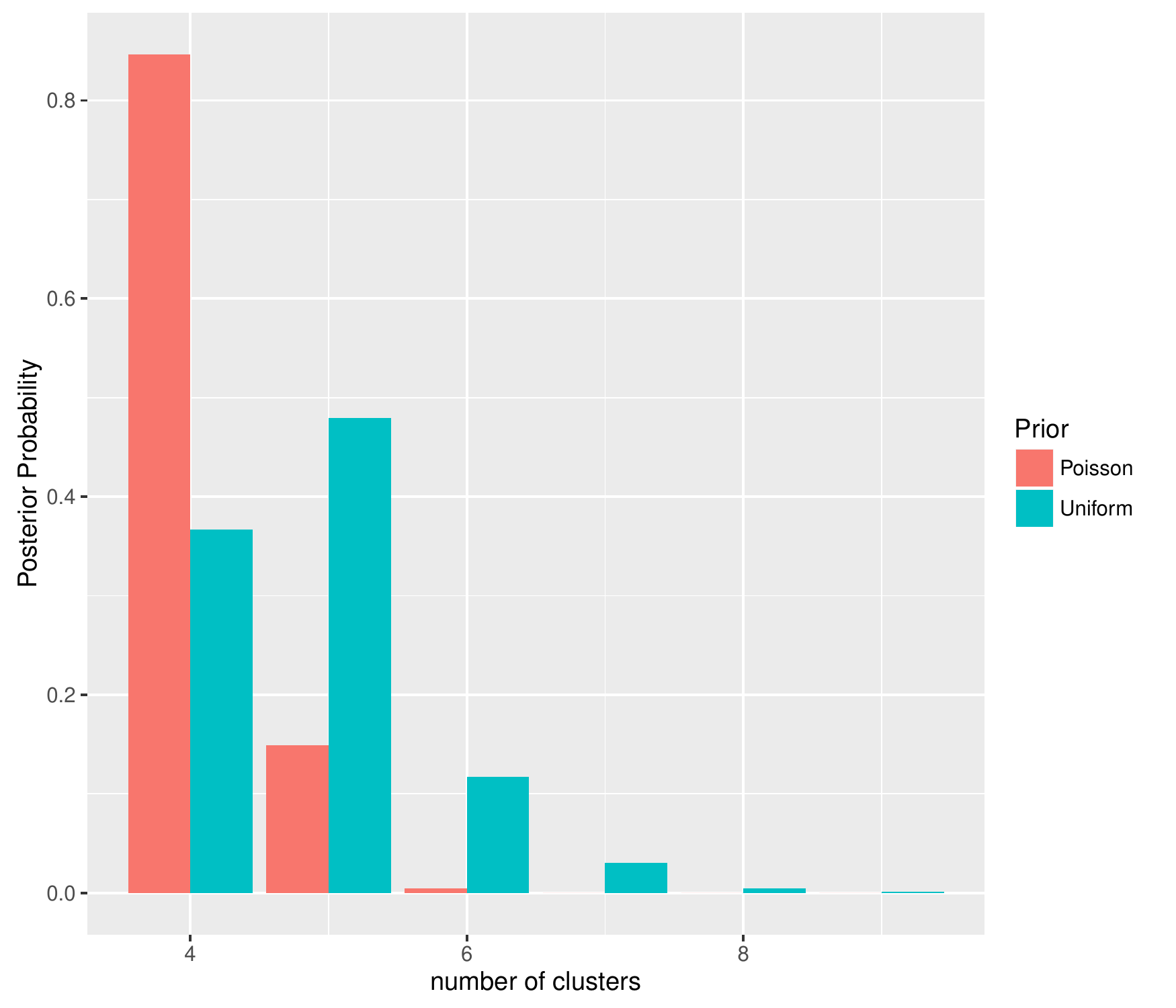}&
\includegraphics[scale=0.35]{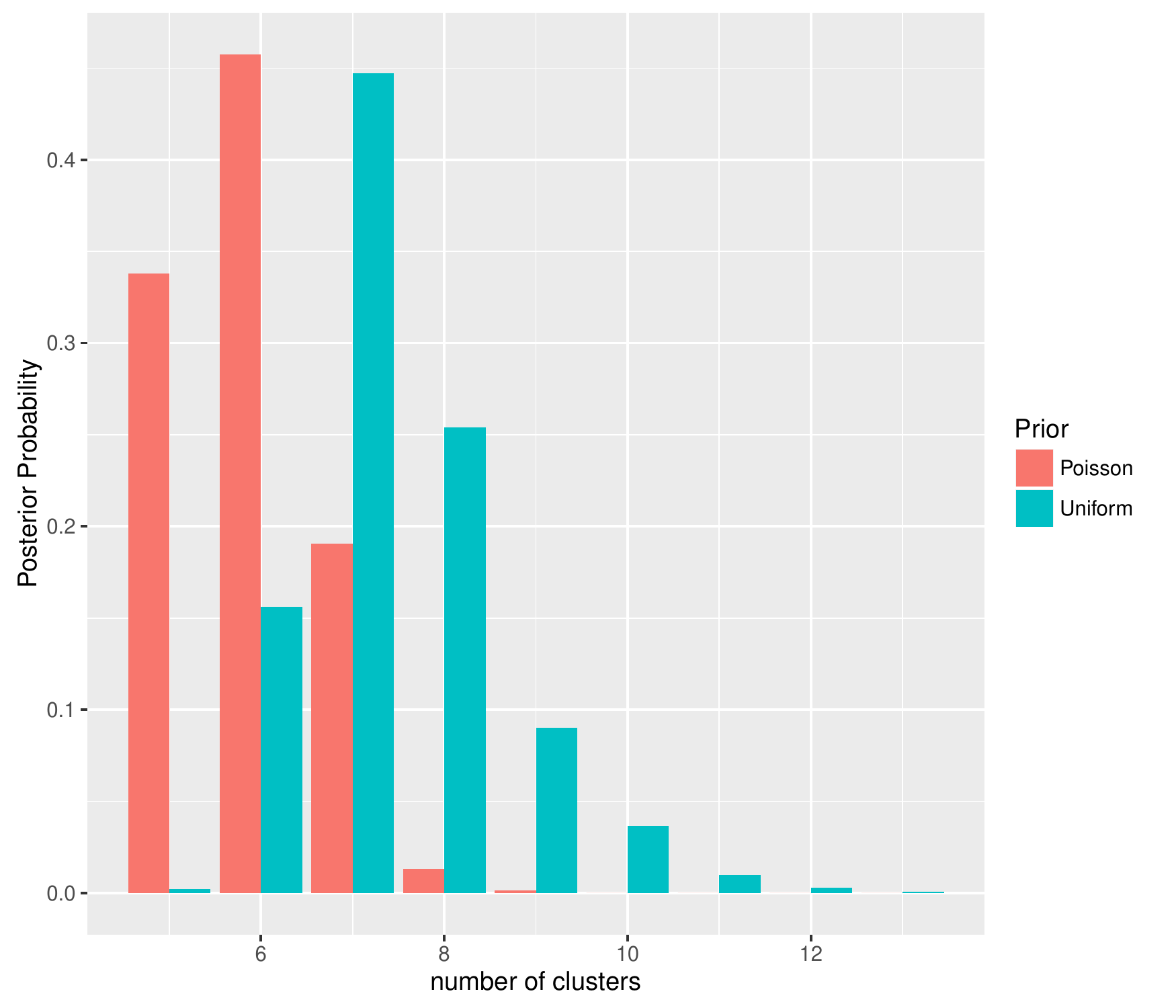}\\
(a) $\alpha = \beta = 1$ (Uniform) & (b) $\alpha = \beta = 0.5$ (Jeffreys)
\end{tabular}
\caption{Zoo dataset: Estimated posterior distribution of the number of clusters when using different parameters $(\alpha,\beta)$ on the Beta prior of $\theta_{kj}$. Both choices of the prior on the number of clusters are considered: truncated Poisson (red) and Uniform (green).}
\label{fig:zoo}
\end{figure}

We test our method considering both prior assumptions on the number of clusters, as well as different hyper-parameters on the prior distribution of $\theta_{kj}$ in Equation \eqref{eq:priorTheta}: we consider $\alpha = \beta = 1$ (default values) as well as $\alpha = \beta = 0.5$. Note that the second choice corresponds to the Jeffreys prior \citep{jeffreys} for a Bernoulli trial. Figure \ref{fig:zoo} displays the estimated posterior distribution of the number of clusters $K$ when $K\in\{1,\ldots,20\}$. This is done with the following commands.

\begin{example}
# read data
> xOriginal <- read.table("zoo.data", sep=",")
> x <- xOriginal[ , -c(1, 14, 18)]
> x <- x[-27, ] # delete 2nd frog
# now transform v14 into six binary variables
> v14 <- xOriginal[-27, 14]
> newV14 <- array(data = 0, dim = c(100, 6))
> for(i in 1:100){
+        if( v14[i] == 0 ){ newV14[i,1] = 1 }
+        if( v14[i] == 2 ){ newV14[i,2] = 1 }
+        if( v14[i] == 4 ){ newV14[i,3] = 1 }
+        if( v14[i] == 5 ){ newV14[i,4] = 1 }
+        if( v14[i] == 6 ){ newV14[i,5] = 1 }
+        if( v14[i] == 8 ){ newV14[i,6] = 1 }
+ }
> x <- as.matrix(cbind(x, newV14))

# apply BayesBinMix using 8 heated chains
> library('BayesBinMix')
> nChains <- 8
> heats <- seq(1,0.6,length = nChains)

# K ~ P{1,...,20}, theta_{kj} ~ Beta(1, 1)
> c1 <- coupledMetropolis(Kmax = 20, nChains = nChains, heats =  heats, binaryData = x,
+       outPrefix = 'poisson-uniform', ClusterPrior = 'poisson', 
+       m = 4400, burn = 400, z.true = z.ground_truth)

# K ~ U{1,...,20}, theta_{kj} ~ Beta(1, 1)
> c2 <- coupledMetropolis(Kmax = 20, nChains = nChains, heats =  heats, binaryData = x,
+       outPrefix = 'uniform-uniform', ClusterPrior = 'uniform', 
+       m = 4400, burn = 400, z.true = z.ground_truth)

# K ~ P{1,...,20}, theta_{kj} ~ Beta(0.5, 0.5)
> c3 <- coupledMetropolis(Kmax = 20, nChains = nChains, heats =  heats, binaryData = x,
+       outPrefix = 'poisson-jeffreys', ClusterPrior = 'poisson', 
+       m = 4400, burn = 400, z.true = z.ground_truth)

# K ~ U{1,...,20}, theta_{kj} ~ Beta(0.5, 0.5)
> c4 <- coupledMetropolis(Kmax = 20, nChains = nChains, heats =  heats, binaryData = x,
+       outPrefix = 'uniform-jeffreys', ClusterPrior = 'uniform', 
+       m = 4400, burn = 400, z.true = z.ground_truth)
\end{example}

\begin{figure}[t]
\includegraphics[scale= 0.45]{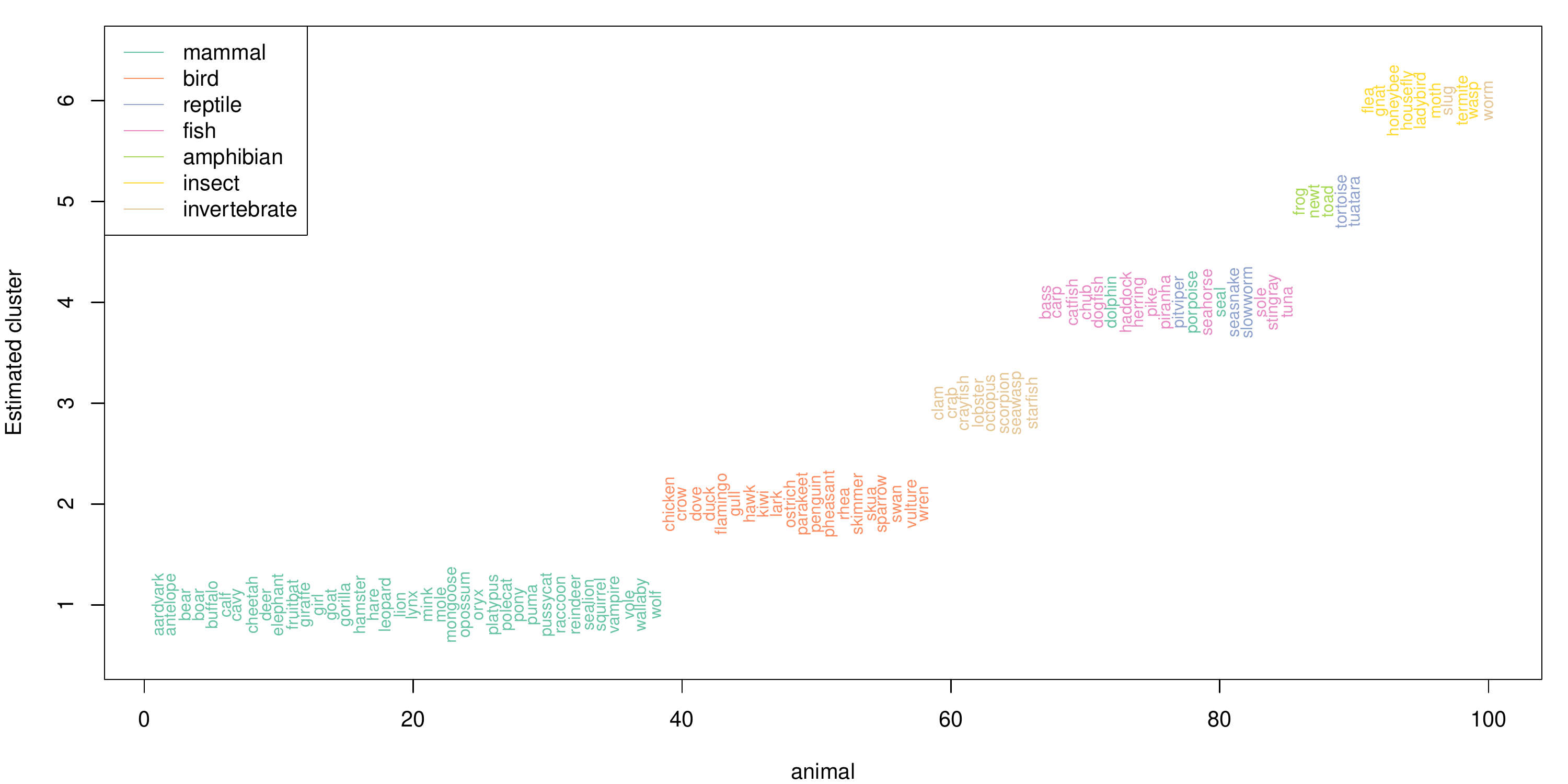}
\caption{Zoo dataset clustering conditionally on the most probable number of clusters $K = 6$ when using the truncated Poisson prior on $K$ and the Jeffreys prior on $\theta_{kj}$. The ground-truth classification of the animals to seven classes is illustrated using different colour for each class.}
\label{fig:zoo_clusters}
\end{figure}

Next, we compare the estimated clusters (for the most probable value of $K$) with the classification of the data into 7 classes (given in the vector \code{z.ground\_truth}). For this reason we provide the rand index (adjusted or not) based on the confusion matrix between the estimated and ground-truth clusters, using the package \CRANpkg{flexclust} \citep{flexclust}.

\begin{example}
> library('flexclust')
> z <- array(data = NA, dim = c(100, 4))
> z[ , 1] <- c1$clusterMembershipPerMethod$ECR
> z[ , 2] <- c2$clusterMembershipPerMethod$ECR
> z[ , 3] <- c3$clusterMembershipPerMethod$ECR
> z[ , 4] <- c4$clusterMembershipPerMethod$ECR
> rand.index <- array(data = NA, dim = c(4, 3))
> rownames(rand.index) <- c('poisson-uniform', 'uniform-uniform',
                                'poisson-jeffreys', 'uniform-jeffreys')
> colnames(rand.index) <- c('K_map', 'rand_index', 'adjusted_rand_index')
> findMode <- function(x){ as.numeric( names(sort(-table(x$K.mcmc)))[1] ) }
> rand.index[ , 1] <- c( findMode(c1), findMode(c2), findMode(c3), findMode(c4) )
> for(i in 1:4){
+       rand.index[i, 2] <- randIndex(table(z[ , i], z.ground_truth), correct = F)
+       rand.index[i, 3] <- randIndex(table(z[ , i], z.ground_truth))
+ }
> rand.index
                 K_map rand_index adjusted_rand_index
poisson-uniform      4  0.9230303           0.7959666
uniform-uniform      5  0.9408081           0.8389208
poisson-jeffreys     6  0.9505051           0.8621216
uniform-jeffreys     7  0.9490909           0.8525556
\end{example}

Note that both rand indices (raw and adjusted) are larger for \code{z[ , 3]}, that is,  the six-component mixture model that corresponds to the Poisson prior on $K$ and the Jeffreys prior on $\theta_{kj}$. A detailed view on the estimated clusters for this particular model is shown in Figure \ref{fig:zoo_clusters}. We conclude that the estimated groups are characterized by animals belonging to the same taxonomy with  very small deviations from the true clusters. Interestingly, in the case that an animal is wrongly assigned to a cluster, notice that the estimated grouping might still make sense: e.g.~ the sea mammals dolphin, porpoise and seal are assigned to the fourth cluster which is mainly occupied by the group 'fish'.

\begin{figure}[t]
\begin{center}
\includegraphics[scale=0.5]{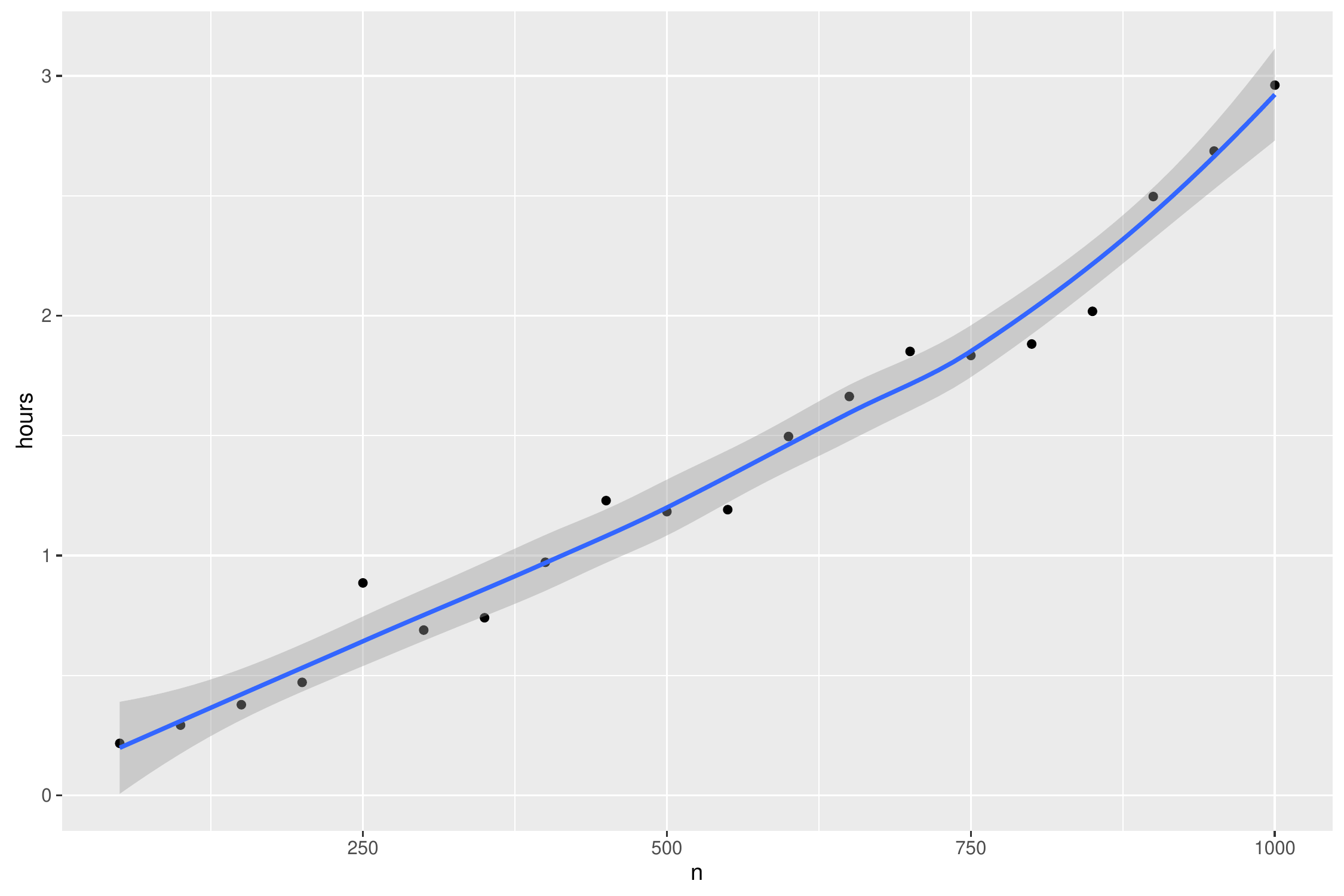}
\end{center}
\caption{Wall clock time of the \code{coupledMetropolis} function when running \code{nChains = 4} heated chains on the same number of parallel threads for \code{m = 1100} cycles, with each cycle consisting of 10 MCMC iterations. At most \code{Kmax = 20} clusters are allowed. For all sample sizes ($n$), the dimension of the multivariate data equals to $d = 100$.}
\label{fig:times}
\end{figure}

\section{Summary and remarks}

The \pkg{BayesBinMix} package for fitting mixtures of Bernoulli distributions with an unknown number of components has been presented. The pipeline consists of a fully Bayesian treatment for the clustering of multivariate binary data: it allows the joint estimation of the number of clusters and model parameters, deals with identifiability issues as well as it produces a rapidly mixing chain. Using a simulation study we concluded that the method outperforms the EM algorithm in terms of estimating the number of clusters and at the same time produces accurate estimates of the underlying model parameters. In the real dataset we explored the flexibility provided by using different prior assumptions and concluded that the estimated clusters are strongly relevant to the natural grouping of the data.

For the prior distribution on the number of clusters our experience suggests that  the truncated Poisson distribution performs better than the uniform (see also \citet{Nobile2007}). Regarding the prior distribution on the Bernoulli parameters we recommend to try both the uniform distribution (default choice) as well as the Jeffreys prior, especially when the sample size is small. An important parameter is the number of heated chains which run in parallel, as well as the temperature of each chain. We suggest to run at least \code{nChains = 4} heated chains. The \code{heat} parameter for each presented example achieved an acceptance ratio of proposed swaps between pairs of chains between $10\%$ and $70\%$. The default choice for the temperature vector is \code{heats = seq(1,0.3,length = nChains)}, however we advise to try different values in case that the swap acceptance ratio is too small (e.g.~  $<2\%$) or too large (e.g.~$>90\%$). Finally, we recommend running the algorithm using at least \code{m = 1100} and \code{burn = 100} for total number of MCMC cycles and burn-in period, respectively. For these particular values of \code{nChains} and \code{m}, Figure \ref{fig:times} displays the wall clock time demanded by the \code{coupledMetropolis} function.

\section{Acknowledgements}

Research was funded by MRC award MR/M02010X/1. The authors would like to thank Rebecca Howard and Dr.~Lijing Lin (University of Manchester) for using the software and reporting bugs to earlier versions. We also thank an anonymous reviewer and Roger Bivand, Editor of the R Journal, for their valuable comments and suggestions that considerably improved the package and the presentation of our findings.

\bibliographystyle{imsart-nameyear.bst}      
\bibliography{main}   

\end{document}